\documentclass{aa}
\usepackage[utf8]{inputenc}
\usepackage{graphicx}
\usepackage{txfonts}
\usepackage{verbatim}
\usepackage{siunitx}
\usepackage{hyperref}
\hypersetup{colorlinks, allcolors=blue}
\usepackage{subcaption}
\usepackage{comment}
\usepackage[labelfont=bf]{caption}
\usepackage{tablefootnote}
\begin{document}

\title{The $\rho$~Oph region revisited with Gaia EDR3}
\subtitle{Two young populations, new members, and old impostors}
\titlerunning{$\rho$ Oph region in Gaia EDR3}
\author{Natalie Grasser\inst{1}
    \and Sebastian Ratzenböck\inst{2}
    \and João Alves\inst{1,2,3}
    \and Josefa Großschedl\inst{1}
    \and Stefan Meingast\inst{1}
    \and Catherine Zucker\inst{4}
    \and Alvaro Hacar\inst{1}
    \and Charles Lada\inst{4}
    \and Alyssa Goodman\inst{3,4}
    \and Marco Lombardi\inst{5}
    \and John C. Forbes\inst{6}
    \and Immanuel M. Bomze \inst{2}
    \and Torsten Möller\inst{2}
}

\institute{
    University of Vienna, Department of Astrophysics, Türkenschanzstrasse 17, 1180 Wien, Austria \\
    \email{natalie.grasser@univie.ac.at}
    \and
    Data Science at University of Vienna, Währinger Straße 29, 1090 Vienna, Austria
    \and
    Radcliffe Institute for Advanced Study, Harvard University, 10 Garden Street, Cambridge, MA 02138, USA 
    \and
    Center for Astrophysics | Harvard \& Smithsonian, 60 Garden St., Cambridge, MA 02138, USA
    \and
    University of Milano, via Celoria, 16, I-20133 Milano, Italy
    \and
    Flatiron Institute, Simons Foundation, 162 Fifth Avenue, New York, NY 10010, USA
}
\date{}

\abstract{Young and embedded stellar populations are important probes of the star formation process. Their properties and the environments they create have the potential to affect the formation of new planets. Paradoxically, we have a better census of nearby embedded young populations than of the slightly more evolved optically visible young populations. The high accuracy measurements and all-sky coverage of \textit{Gaia} data are about to change this situation.}{This work aims to construct the most complete sample to date of young stellar objects (YSOs) in the $\rho$~Oph region.}{We compile a catalog of 1114 Ophiuchus YSOs from the literature and cross-match it with the \textit{Gaia} EDR3, \textit{Gaia}-ESO, and APOGEE-2 surveys. We apply a multivariate classification algorithm to this catalog to identify new, co-moving population candidates.}{We find 191 new high-fidelity YSO candidates in the \textit{Gaia} EDR3 catalog belonging to the $\rho$~Oph region. The new sources appear to be mainly Class\,III M stars and substellar objects and are less extincted than the known members, while we find that 28 of the previously unknown sources are YSOs with circumstellar disks (Class\,I or Class\,II). The analysis of the proper motion distribution of the entire sample reveals a well-defined bimodality, implying two distinct populations sharing a similar 3D volume. The first population comprises young stars' clusters around the $\rho$~Ophiuchi star and the main Ophiuchus clouds (L1688, L1689, L1709). In contrast, the second population is slightly older ($\sim 10$ Myr), more dispersed, has a distinct proper motion, and is possibly from the Upper Sco group. The two populations are moving away from each other at about 4.1\,km/s and will no longer overlap in about 4\,Myr. Finally, we flag 17 sources in the literature sample as likely impostors, which are sources that exhibit large deviations from the average properties of the $\rho$~Oph population. Our results show the importance of accurate 3D space and motion information for improved stellar population analysis.}{}

\maketitle


\section{Introduction}

Since the development of millimeter-wave receivers and infrared (IR) detectors in the 1970s, local star formation studies have mostly concentrated on the densest star-forming structures in molecular clouds. Successive generations of instruments have opened a fundamental window into molecular cloud structure, cloud fragmentation, and collapse and have unveiled the dust-enshrouded young stellar object (YSO) populations in nearby clouds. This approach has generated an almost paradoxical situation where we currently know more about the very young dust-obscured populations than we know about the more evolved and optically revealed population in nearby star-forming regions. 

More evolved YSOs show less IR excess emission and escape detection in IR surveys but are critical to reconstructing a region's star formation history. Therefore, identifying the young optically visible population is essential for reconstructing a star formation event. Moreover, sources in the unobscured environments of nearby star-forming gas include some of the lowest-mass objects (brown dwarfs and planetary-mass objects) and some of the closest proto-planetary disks we can study, the latter becoming important targets for resolved ALMA studies (e.g., \citealt{ALMA2015}) in the submillimeter wavelength range.

Optical data from the \textit{Gaia} mission \citep{GaiaMission2016}, with its exquisite sensitivity and all-sky coverage, have changed this situation. With its latest data release, the mission has made a breakthrough in terms of studies of gas shape and motion \citep{Grossschedl2018, Grossschedl2020} and previously unknown young stellar structures \citep{Meingast2019, Meingast2020}, significantly improving upon its second data release, \textit{Gaia} DR2 \citep{GaiaDR22018}. In this work, we revisit one of the nearest star-forming regions, the $\rho$~Ophiuchi region, by using the newly available \textit{Gaia} Early Data Release 3 (EDR3) data \citep{GaiaEDR32020}.

The $\rho$~Ophiuchi ($\rho$ Oph) star-forming region \citep{Wilking} is one of the nearest active star-forming regions, at a distance of approximately 139 pc \citep{Lombardi2008,Zucker}. It comprises the cluster of young stars around the  $\rho$\,Ophiuchi star \citep{Pillitteri2016} and the young stars associated with the dense gas in the Ophiuchus cloud complex, mainly the L1688, L1689, and L1709 clouds \citep{Loren1989a, Loren1989b}. Due to its youth and proximity to Earth, it has played an essential role in many star formation studies, in particular in the definition of the YSO classes \citep{WilkingLada,Lada1984,Andre,Greene}. The $\rho$~Oph region is located in the foreground of the southeastern edge of Upper Scorpius, which is a subgroup of the Scorpius-Centaurus OB association, and has a distance of around 145 pc \citep{Wilkinson2018-rx}. It has long been suspected that star formation in the $\rho$~Oph region was triggered by feedback from massive stars from Upper Sco \citep{Vrba1977AJ, LorenWootten1986, Loren1989a, Loren1989b, deGeus1992}.

The youngest stars in the region are associated with the densest gas in the Ophiuchus cloud complex, mostly L1688, with an average age of about 0.3 Myr \citep{GreeneMeyer1995, LuhmanRieke1999}, while the stellar population on the lower column density surface has an average estimated age of 2 to 5 Myr \citep{Wilking,Erickson2011-vu}. There are three main dark clouds in the $\rho$~Oph complex, the mentioned Lynds dark clouds L1688, L1689 and L1709 \citep{Lynds1962, Loren1989a, Loren1989b}. The large column density toward particular regions in these clouds, where the optical extinction can reach values of up to A$_\text{V}$ above 40--50\,mag \citep{WilkingLada, Wilking1989, Lombardi2008}, make IR observations essential for studying the embedded young stellar population in the cloud. There is a rich embedded cluster of YSOs in L1688, which is mostly invisible at optical wavelengths, whose stars have not yet dispersed \citep{Ducourant}. 

In this paper, we apply the recently developed method from \citet{Ratzenboeck} to \textit{Gaia} EDR3 data, to unveil the most complete sample to date of YSOs toward the $\rho$~Oph region. The method uses the astrometric properties of known YSOs in combination with a bagging classifier of one-class support vector machines (OCSVMs) on \textit{Gaia} EDR3 data to perform a 5D search (3D positions and 2D proper motions) for possible new population members. The algorithm creates a hyper-surface around the positional and proper motion distribution of the input samples in a 5D space to find new sources with similar properties. Radial velocities of the input population are also necessary for constraining the models. We remove models that identify stars with significantly different 3D velocities than the those of training set. 

In Sect.~\ref{sec:data} we present the data used in this work; this includes known sources from the literature, which we cross-matched with further astronomical surveys. In Sect.~\ref{sec:methods} we summarize how the classification algorithm operates to identify new sources. We present the results of the algorithm in Sect.~\ref{sec:results}, including a detailed analysis. In Sect.~\ref{sec:discussion} we discuss some implications of our findings.
 
\section{Data} \label{sec:data}

\subsection{Literature catalog}

In this section, we summarize how we compiled our literature catalog of $\rho$~Oph sources. This work is based on studies of $\rho$~Oph and L1688 from 11 papers, which are summarized in Table \ref{tab:literature}, including the number of sources utilized from each work, which results in a total of 1114 sources. We note that the same source can be presented in more than one work. We assign each paper a digit for citation purposes in our final catalog. Some papers also include sources from IR observations, which are essential for a complete sample due to the high optical extinction in the region and for identifying Class\,II and earlier Class YSOs. The highest number of sources are provided by \cite{Wilking}, \cite{Canovas} and \cite{Esplin}. Duplicates were removed with an internal match within a 1.0 arcsec match radius and an internal match on the \textit{Gaia} source IDs. Our result is a final literature table of 1114 unique sources.

\begin{table*}[t!]
    \caption{Overview of the literature that was used to collect young stellar members of the $\rho$~Oph region.}
    \centering
    \begin{tabular}{llrr}
    \hline
    \hline
    Paper & Method & Sources used & Ref \\
    \hline
    \cite{Greene} & mid-IR photometric study & 56 & 1\\
    \cite{Haisch} & near- and mid-IR observations & 13 & 2\\
    \cite{Padgett} & Multiband Imaging Photometer for \textit{Spitzer} (MIPS) point-sources & 46 &3 \\
    \cite{Wilking} & X-ray and IR photometric and spectroscopic surveys & 316 & 4\\
    \cite{Evans} & \textit{Spitzer} c2d Legacy survey & 292 & 5 \\
    \cite{Dunham} & \textit{Spitzer} c2d and GB Legacy surveys & 292 & 6\\
    \cite{Rigliaco} & dynamical analysis with \textit{Gaia}-ESO survey & 45 & 7 \\    
    \cite{Ducourant} & near-IR observations to determine proper motions & 82 & 8\\
    \cite{Canovas} & density-based clustering algorithms with \textit{Gaia} DR2 & 831 & 9\\
    \cite{Sullivan} & radial velocity survey with data from IR spectrographs & 34 & 10\\
    \cite{Esplin} & astrometry from \textit{Gaia} DR2, proper motions from \textit{Spitzer} & 373 & 11 \\
    \hline
    \end{tabular}
    \tablefoot{The table lists the used methods and the number of sources we obtained from each paper, resulting in a total of 1114 literature sources. We note that the same source can be presented in more than one work.}
    \label{tab:literature}
\end{table*}

\citealt{Sullivan} provide radial velocities on their sources, while \cite{Ducourant} provide proper motions on their sources. Astrometric data (proper motions, parallaxes, and radial velocities) for the remaining sources were obtained by selecting three surveys for cross-matching with our literature sample, which is essential for identifying new sources with the algorithm. The \textit{Gaia} survey provides us with unprecedented astrometry with improved quality and statistics compared to any previous comparable survey, such as \textsc{Hipparcos} \citep{Hipparcos1997}. Therefore, proper motions and parallaxes were obtained from \textit{Gaia} EDR3 \citep{GaiaEDR32020}. To complement \textit{Gaia} astrometry and constrain the models of the algorithm, we combined it with radial velocities from APOGEE-2 \citep{APOGEE}, a large-scale spectroscopic survey conducted in the near-infrared, and \textit{Gaia}-ESO \citep{GES}, a spectroscopic survey by the European Southern Observatory (ESO) combined with the \textit{Gaia} astrometry catalog. Radial velocities from these surveys deliver superior resolution and statistics compared to radial velocities from \textit{Gaia}.

A cross-match of the literature sources with data from \textit{Gaia} EDR3 yielded a total of 675 matches, which is 60.5$\,\%$ of the entire literature sample, leaving many sources without \textit{Gaia} equivalents. One explanation for this is that \textit{Gaia} is only sensitive to optical wavelengths, while many of the obtained literature sources are too embedded in the cloud and can only be observed at IR wavelengths. Additionally, several sources, such as from \cite{Esplin}, are brown dwarfs, which are often too faint to be seen by \textit{Gaia}. A cross-match of the total literature sources with APOGEE-2 resulted in 188 matches, while a cross-match with \textit{Gaia}-ESO data yielded 61 matches in our literature catalog. For sources with multiple measurements, higher priority was given to surveys with higher accuracy. Therefore we use \textit{Gaia} proper motions and parallaxes over those obtained from the literature. For sources with multiple radial velocity values, data from \textit{Gaia}-ESO has the highest priority, followed by APOGEE-2 and then \textit{Gaia}.

The distances to the sources were calculated through the inverse of the parallax, which is a good approximation for the relatively close distance to the region of about 130--140\,pc (e.g., \citealt{Luri2018}). Furthermore, the tangential velocities $v_{\alpha}$ and $v_{\delta}$, as well as their errors, were calculated through the proper motions and parallaxes, as shown in Equations \ref{equ:valpha}--\ref{equ:vdelta}. For a better overview, we list the symbols and abbreviations of frequent parameters used throughout this paper:

\begin{itemize}
\item $\alpha, \delta$ (deg): right ascension and declination
\item $l, b$ (deg): galactic longitude and latitude
\item $\varpi$ (mas): parallax of the sources
\item $d$ (pc): distance to the sources, inverse of parallax
\item $\mu_{\alpha}^{*}$ (mas/yr): $\mu_{\alpha}$cos($\delta$), proper motion along $\alpha$
\item $\mu_{\delta}$ (mas/yr): proper motion along $\delta$
\item $v_r$ (km/s): heliocentric radial velocity
\item $v_{\alpha}, v_{\delta}$ (km/s): tangential velocities along $\alpha$ and $\delta$
\item $v_{l}, v_{b}$ (km/s): tangential velocities along $l$ and $b$
\item $X,Y,Z$ (pc): positions in Galactic Cartesian coordinates, where $X$, $Y$, and $Z$ point toward the Galactic center, the direction of the Galactic rotation, and the north Galactic pole, respectively
\item $U,V,W$ (km/s): velocities in Galactic Cartesian coordinates
\end{itemize}

\subsection{Impostors} \label{sec:imp}

We have discovered several sources within the literature catalog that have properties that do not fit very well to the region's average astrometric values. In Appendix~\ref{app:train} we list the interval ranges in which most of the distance, radial velocity, and tangential velocity values in $\rho$~Oph are found, which were used to create a training set (Sect.~\ref{sec:training-set}). There are 28 sources that have at least one of these values outside our defined intervals and smaller errors than the upper limits listed in Appendix~\ref{app:train}. However, some of them have values that are still close to the interval limits and could therefore still be a part of $\rho$~Oph, since deviating motions can be caused by interactions in the cluster or by multiple stellar systems. There are, nonetheless, several sources with very large radial velocity deviations from the average. Therefore, we identified all sources with radial velocities $v_r<-30$ and $v_r>20$ and errors $< 3$ as uncertain members and labeled them as impostor candidates in our catalog. We found 17 of such impostor candidates among the literature sources. However, it is important to note that these deviating radial velocities could be caused by multiplicity, such as binary star systems, and could therefore still be members. Due to this uncertainty, and since our intervals are more or less arbitrarily defined, we chose not to remove these impostors from our catalog. Instead, we created a separate column named ``Impostors,'' where they are labeled with a ``1'' and all others are labeled with a ``0.''

\section{Methods} \label{sec:methods}

In our work, we applied the classification strategy described in \citet{Ratzenboeck} for identifying new members of the $\rho$~Oph region in the \textit{Gaia} EDR3 catalog. The goal of \cite{Ratzenboeck} was to model the extent of the Meingast~1 stellar stream \citep{Meingast2019} in the combined space of proper motions and positions and subsequently use it to identify new members in \textit{Gaia} DR2, while we use the latest data release EDR3. The model consists of multiple OCSVM classifiers in a bagging ensemble. In the following we refer to sources classified by the OCSVM as members of a stellar population as ``predicted'' members. Based on the model quality, the prediction set contains known and potentially new candidate sources. In the following we discuss means of selecting high quality models via prior assumption filters.


\subsection{Training set selection} \label{sec:training-set}
To provide reliable sources for the classification algorithm, we created a training set by removing outliers and applying quality cuts. The quality cuts are described in Appendix~\ref{app:train}, where we also present the training set. To guarantee a high-fidelity training set, we limited our selection to sources with radial velocity measurements. Since the hypersurface created by the OCSVM algorithm depends heavily on the distribution of peripheral sources, it is susceptible to outliers. The use of a soft-margin SVM somewhat mitigates this, but to further reduce the effect of potential contaminants on the final model shape, we removed the most extreme outliers from the training set as well. To do so, we estimated the local outlier factor \citep{Breunig2000} of each source in 5D and removed 5$\,\%$ of the training set with the highest outlier factor. This removal lead to a final training set of 150 sources, which corresponds to 13.5$\,\%$ of the literature sample.

\subsection{Model selection and prior assumptions}
Due to the high model flexibility of OCSVMs, choosing adequate model parameters is critical to guarantee a suitable description of the stellar system. Instead of directly selecting models in the OCSVM hyperparameter space, \cite{Ratzenboeck} have suggested to constrain the models via prior assumptions they have to adhere to, implicitly tuning the model parameters. In addition, as summary statistics, prior assumptions are usually much easier to interpret compared to the original OCSVM parameters. Each set of prior beliefs corresponds to a distribution of allowed models in the input parameter space, such that there is a mapping from a prior assumption tuple to regions in the OCSVM parameter space that contain models that adhere to the given rule set. Instead of explicitly characterizing this map, we sampled uniformly from the OCSVM hyperparameter space and removed unfit models. To determine a set of prior assumptions for identifying new high-fidelity $\rho$~Oph members, we considered their application in \cite{Ratzenboeck}. The prior assumptions were motivated by the training set selection process. Since only sources with radial velocities were previously identified to be part of the Meingast~1 stream, the authors formulated prior assumptions based on completeness arguments regarding radial velocities. Specifically, the goal was to find still unknown members without radial velocity measurements, which were confined to the training set extent. However, the $\rho$~Oph training set selection function is much more complex as we combined radial velocity information across multiple data surveys. This also means we have much less information about potentially concealed $\rho$~Oph members. Therefore, we adjusted the previous assumptions to the $\rho$~Oph population. In the following, we briefly discuss the selection of the six prior assumptions constraining models via the number and distribution of predicted sources.




\subsubsection{Population size} \label{sec:model_pa}
Firstly, we aim to restrict the number of sources a model identifies. Because the $\rho$~Oph population has been studied extensively --- with some studies using \textit{Gaia}  data as well --- we do not expect to find a dramatic increase in overall population size. Based on the number of \textit{Gaia} EDR3 sources in the literature catalog, we estimated a very conservative upper limit of a maximum population size of about twice the number of sources from the literature catalog that have \textit{Gaia} source IDs to be predicted by a single model, setting it to $1400$ maximal members. We note here that the prior assumption restrictions only apply to single models, meaning the model ensemble, as a final classifier can exceed individual or multiple prior assumption limits.

\subsubsection{Contamination fraction}
Secondly, we constrained the contamination fraction of predicted sources across models. The contamination fraction is determined via the 3D velocity distribution of $\rho$~Oph candidate sources. Precisely, we first modeled the 3D velocity distribution of the training samples as a single Gaussian distribution. The mean and covariance matrix were determined by maximizing the likelihood of the training data. Subsequently, we defined the contamination as the fraction of sources outside the 3$\sigma$ (99.7$\,\%$) range of the training set. In practice, we observe very few radial velocities in the predicted set for a single model, and, therefore, the contamination fraction assumption has a minor effect for removing single models. This effect is highlighted in Fig.~\ref{fig:vr_contamination}, where we see an almost constant and maximal number of models adhering to the contamination rule for various maximal values. Since the influence is small across such a large range, we set it to a value of 15$\,\%$. 

\subsubsection{Estimated extent and systematic shift}
Lastly, we want to constrain the extent of predicted $\rho$~Oph members in position and proper motion space. This was done by measuring the dispersion and systematic shift between training and predicted member distributions. We characterize the dispersion in position and proper motion space by the mean deviation of its member stars to their centroid. The prior assumption corresponds to a constraint on the ratio between the average predicted deviation to the average training deviation. For further details, we refer to Appendix B in \cite{Ratzenboeck}. In the case of $\rho$ Oph, we cannot give a concrete estimate on the expected extent of unknown members in position and proper motion space. Instead, we motivate a range of maximal values. We postulated a constraint on the parameter to be within 1, which constrains the predicted extent to the training set extent, and 2, where models can have twice the dispersion of the training set. We explicitly separated the positional from the proper motion axes since both dispersion measures have physically different meanings, and we might want to restrict one more than the other. 

To avoid systematic shifts of the predicted to the training set distribution, we constrained the distance between the centroids of the training and predicted sources. We measured the centroid distance in terms of the mean deviation of the training set sources. A value of one would correspond to a centroid shift with a distance of one mean deviation from the training centroid. Again, finding a precise value is not straightforward, as the value cannot be properly inferred for the unknown $\rho$~Oph population. Therefore, we limited the maximum shift parameter to a range between 0.1 and 0.7, which we consider already a quite large systematic deviation from the training set.

\subsection{Building the $\rho$ Oph classifier} \label{sec:ophclass}

We subsequently searched for model ensembles within these three parameters, the mean deviation in position, proper motions, and the maximal systematic shift, while keeping the other two prior assumptions, the maximum number of predicted sources, and the maximum contamination fraction, fixed. As stated in Sect. \ref{sec:model_pa}, a prior assumption tuple corresponds to a model ensemble that adheres to the respective beliefs. For each of these ensembles, we determined a stability threshold by minimizing the Kullback–Leibler (KL) divergence \citep{kullback1951} between the 3D velocity distributions of training and predicted $\rho$~Oph members (see Appendix~\ref{app:stab} for more details). We randomly selected 100 prior assumption tuples within their respective range, resulting in 100 model ensembles with a corresponding stability threshold. 

To select single or multiple suitable classifiers from this space of model ensembles, we considered the following. We aimed to maximize the number of predicted $\rho$~Oph sources while minimizing the number of contaminants in our final prediction set. Thus, we studied the distribution of the number of predicted sources over the contamination fraction across the 100 model ensembles. The contamination fraction is determined via the ratio of predicted sources outside the $3 \sigma$ range of the training velocity distribution. The distribution of the 100 randomly sampled model ensembles can be seen in Figure \ref{fig:pa_sampled}. We observed a clear trend for high-contamination models, which tend to have larger velocity dispersion and interestingly a rather low systematic shift to ``good'' models. This sample of low-contamination models were identifying possibly new $\rho$~Oph members in a nonsymmetric region around the training set. To construct the final classifier, we combined the predictions of the 90 models with the lowest contamination fraction of < 28$\,\%$\footnote{The contamination fraction is determined without any quality filters applied.}, corresponding to the two left-most columns of models in the top row of Figure~\ref{fig:pa_sampled}. Finally, we determined a stability threshold for the final ensemble following the procedure outlined in Appendix~\ref{app:stab}. Doing so, we obtained a stability threshold of $4\,\%$. To properly validate the final classifier, we had to consider the previously untouched information, the distribution of sources in the Hertzsprung-Russell Diagram (HRD). In order for the predicted sources to be actual members of the $\rho$~Oph population, they must follow the same isochrone as the training set. Therefore, we determined the residuals of predicted sources to the best fitting isochrone on the training set, where we obtained an age of about 5~Myr, and compared them to the training set residuals. In Figure~\ref{fig:stab_isochrone}, the standard deviation of the training set residuals and predicted residuals can be seen, highlighting an almost perfect agreement with the training data across the full stability range.

\section{Results} \label{sec:results}

In this section we present the results of the algorithm. Sources from the literature are labeled as ``Known'' while the new sources are labeled as ``New.'' The following plots in this section show the known sources in blue, the new sources in red, and a control sample in gray, labeled as ``Control,'' which serves as a comparison. The control sample was selected in a relatively dust-free region to the Galactic west of $\rho$~Oph at the same galactic latitude, within $346^{\circ}\leq l \leq 349^{\circ}$ and  $15^{\circ}\leq b \leq 18^{\circ}$.

\subsection{Predicted sources}

A total of 791 sources in the \textit{Gaia} EDR3 catalog were predicted by the algorithm as belonging to the $\rho$~Oph region, based on the properties of the training set. The predicted sources include a total of 229 new sources that are not in the $\rho$~Oph literature catalog. A total of 562 of the predicted sources are already part of the literature sample of 1114 known sources, meaning that 50.4\,\% of the literature sources were recovered by the algorithm.

\begin{figure}[t!]
    \centering
    \includegraphics[width=\columnwidth]{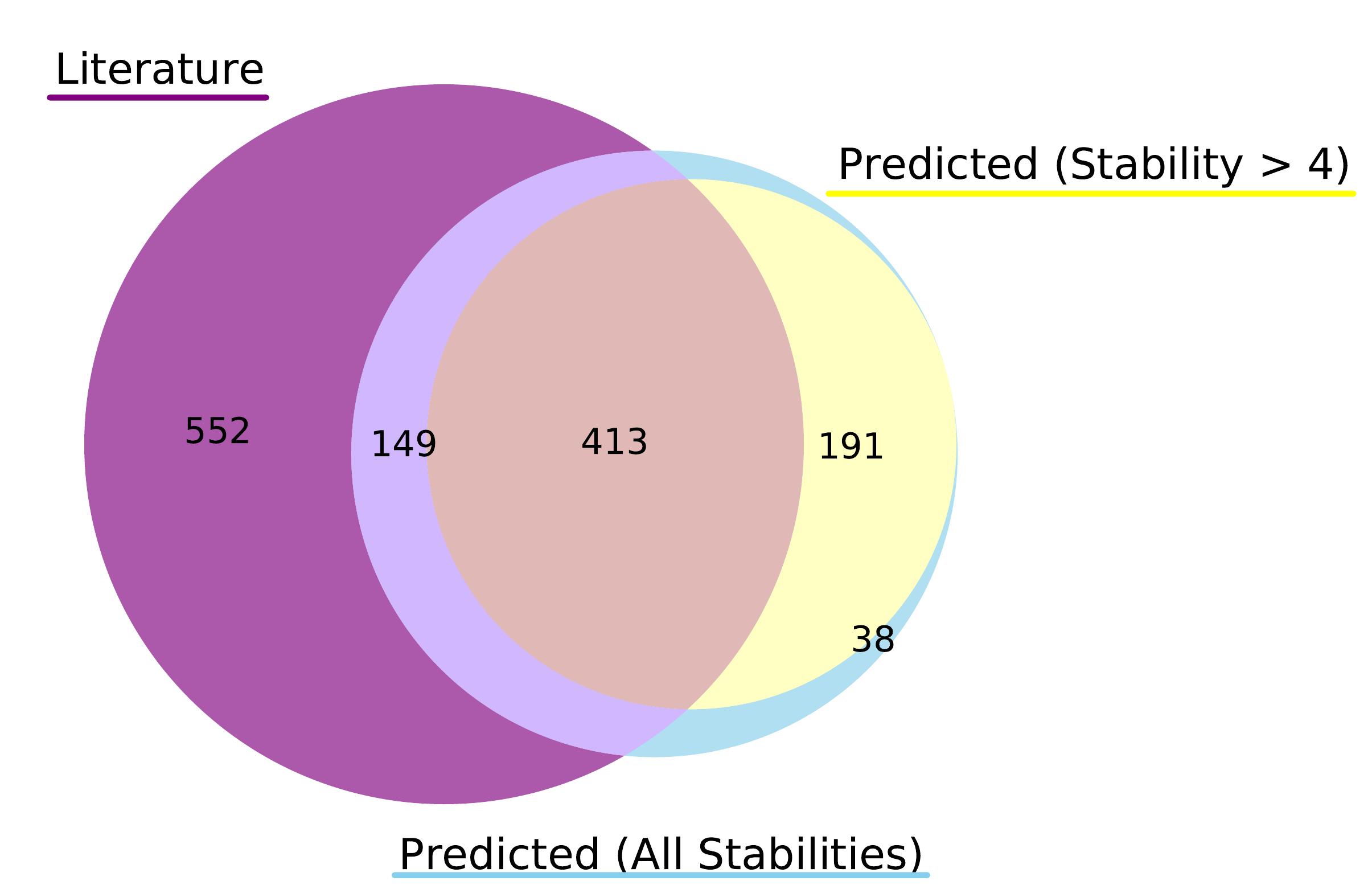}
    \caption{Venn diagram depicting the amount of sources in the literature sample, the predicted sample, and the amount of sources both of them have in common. In total, 791 sources were predicted by the algorithm. Of these, 229 are new sources, with 191 having a stability > 4. 562 of the total predicted sources are already among the 1114 literature sources, 413 of those having a stability > 4. 552 literature sources were not predicted by the algorithm.}
    \label{fig:venn}
\end{figure}

Only the sources with stability > 4\,\%, namely 191 of the new sources, are considered in the following results. These new sources together with the known ones result in 1305 total sources in the $\rho$~Oph region, while when excluding impostor sources we end up with 1288 high probability members. In our final catalog, we also include the new sources predicted by the algorithm with a stability < 4\,\%, resulting in a table of 1343 total sources. Figure~\ref{fig:venn} visualizes the amount of shared sources in the literature and the prediction set in a Venn diagram, showing sources with a prediction stability > 4\,\% and all stabilities. More information on the stability can be found in Appendix~\ref{app:stab}. 
An overview of the final numbers of sources per (sub)sample is given in Table~\ref{tab:overview-numbers}.
A column overview of the final master catalog of the $\rho$~Oph young stellar members is presented in Appendix~\ref{app:tab}.

A probable reason for the relatively small overlap in Figure~\ref{fig:venn} (the algorithm only predicts 50.4$\,\%$ of the known sources from the literature) is the fact that many of the literature sources were obtained through IR surveys since embedded stars in $\rho$~Oph cannot be detected at optical wavelengths. Furthermore, some sources in the literature are impostors, as described in Sect.~\ref{sec:imp}. It is also important to note that only 675 literature sources (60.6$\,\%$) have \textit{Gaia} EDR3 IDs. Therefore, the algorithm has effectively recovered 83.3$\,\%$ of literature sources that are in \textit{Gaia} EDR3. \textit{Gaia} is an optical telescope, hence it is insensitive to high extinction sources in the L1688 dense clump, where the peak of the surface density of YSOs in the cloud complex is located (e.g., \citealt{OritzLeon2017,Ducourant}). Sources not visible at optical wavelengths cannot be predicted by the algorithm.

\subsection{Astrometric properties} \label{subsec:ast}

\begin{figure}[t!]
    \centering
    \includegraphics[width=\columnwidth]{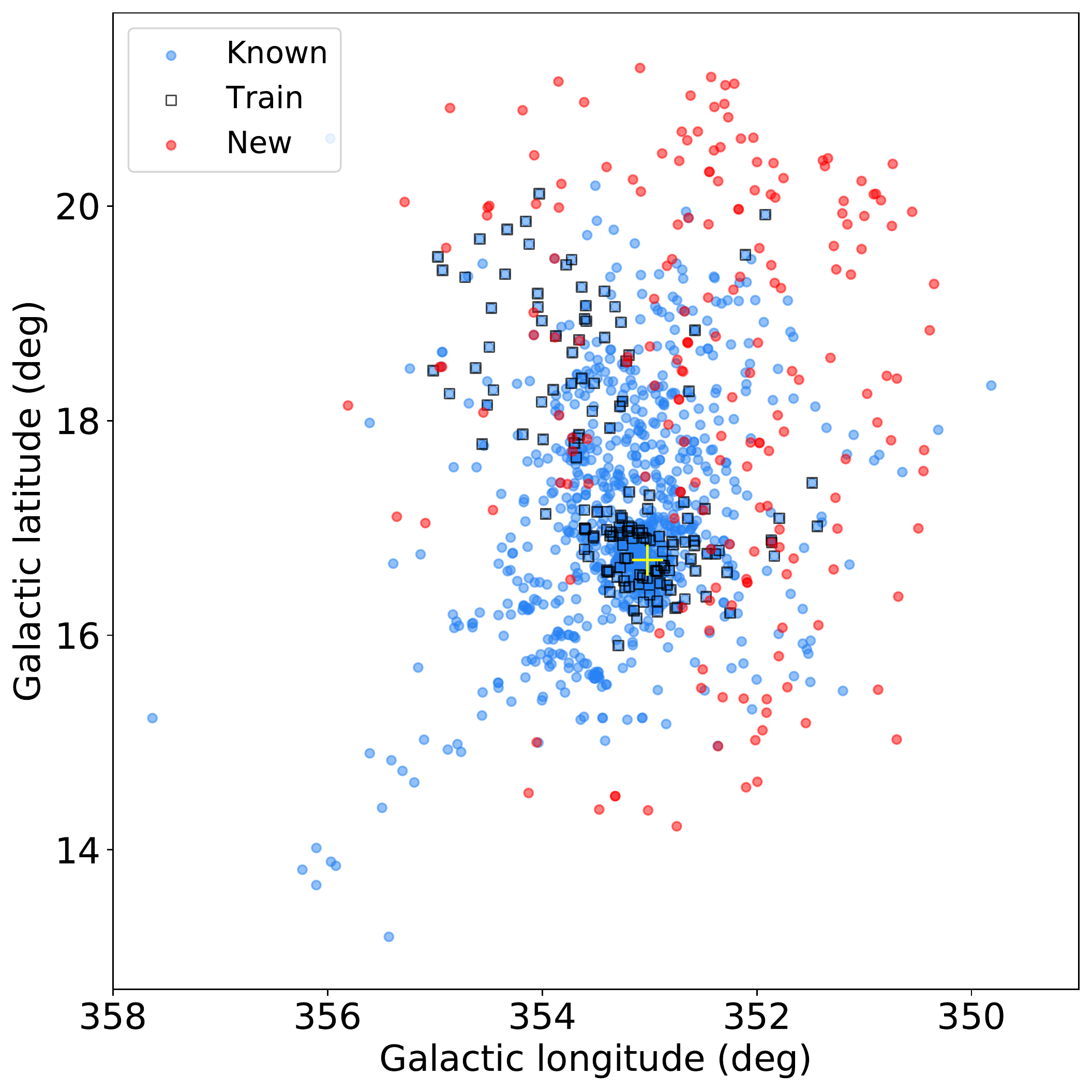}
    \caption{Distribution of $\rho$~Oph sources in galactic coordinates. The known sources from the literature are in blue, while new sources are in red. Sources from the training set are represented by black squares. The approximate location of the extinction peak is marked by a yellow cross.}
    \label{fig:sky}
\end{figure}

Figure \ref{fig:sky} shows the distribution of the $\rho$~Oph sources in galactic coordinates with the known sources in blue and the new ones in red. The 150 sources in the training set, labeled as ``Train,'' are included as unfilled black squares for comparison. As can be seen in the figure, the new sources are more dispersed, with many of them being shifted toward the Galactic north, west, and south of the known sources. Hardly any new sources were found near the core of the cloud and toward the Galactic east. The extinction peak of the L1688 cloud, marked by a yellow cross in the figure, lies at around $l\approx 353.0^{\circ}$ and $b\approx 16.7^{\circ}$ (Alves et al., in prep). It is most likely responsible for the lack of new sources in the core since sources with a high optical extinction cannot be detected by \textit{Gaia}. Furthermore, the core region is the most thoroughly studied part of $\rho$~Oph by previous surveys, thus it is unsurprising that few new sources were found near the core region.

\begin{figure}[t!]
    \centering
    \includegraphics[width=\columnwidth]{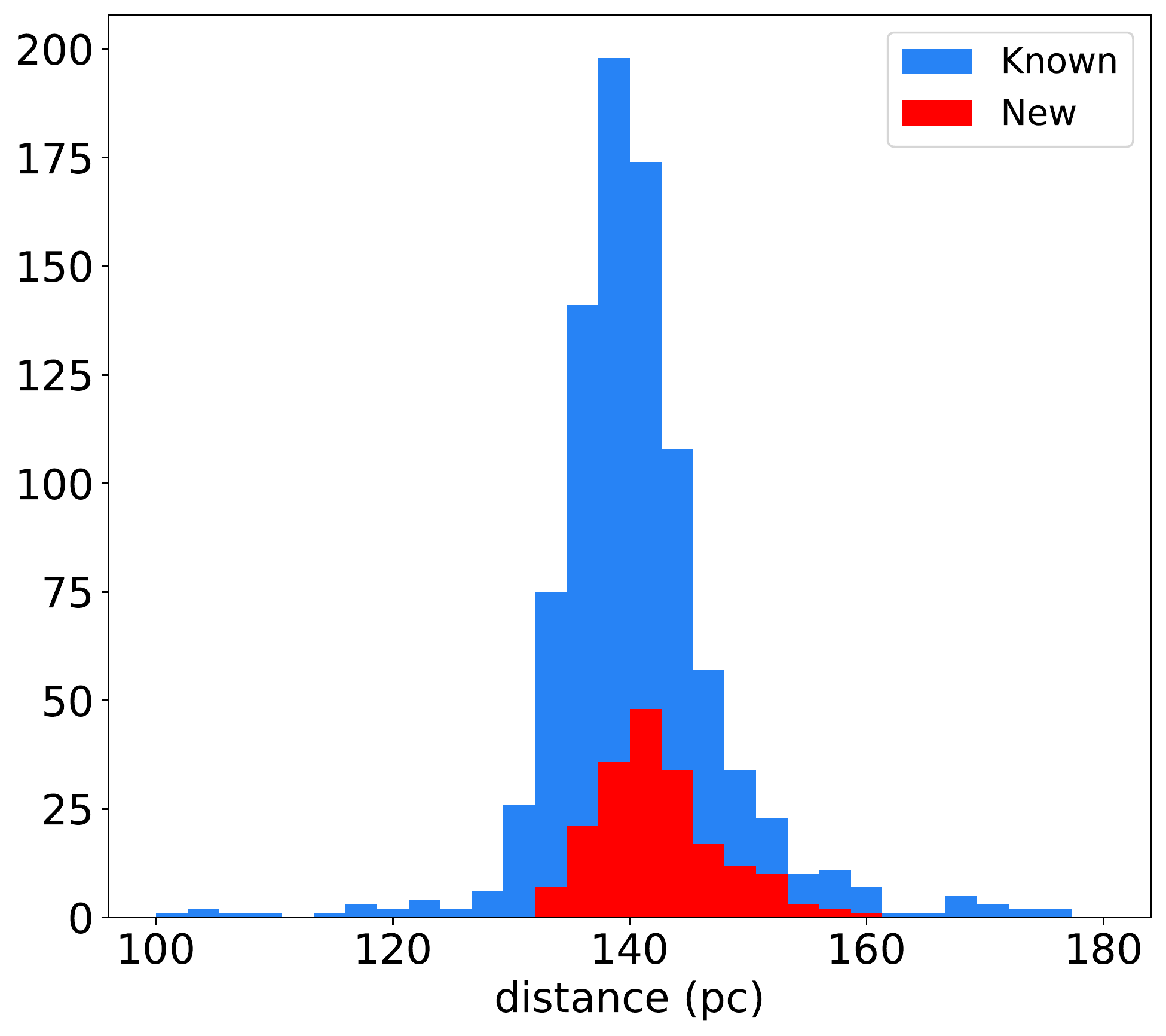}
    \caption{Histogram of distances to the $\rho$~Oph sources. The distribution of the known sources from the literature is in blue, and the new sources are in red.}
    \label{fig:dist}
\end{figure}

Figure~\ref{fig:dist} shows a histogram of the distances to the $\rho$~Oph sources, which were determined through the inverse of their parallaxes. Most of the sources are clustered around a mean distance of approximately 140 pc (see Table~\ref{tab:avg}), which agrees well with the literature value of around 139 pc \citep{Zucker}. In general, the average astrometric properties of the known and new sources are very similar and overlap within $\pm 1\sigma$ (see Table~\ref{tab:avg}), further confirming that they belong to the same region.

\begin{figure}[t!]
    \centering
    \includegraphics[width=\columnwidth]{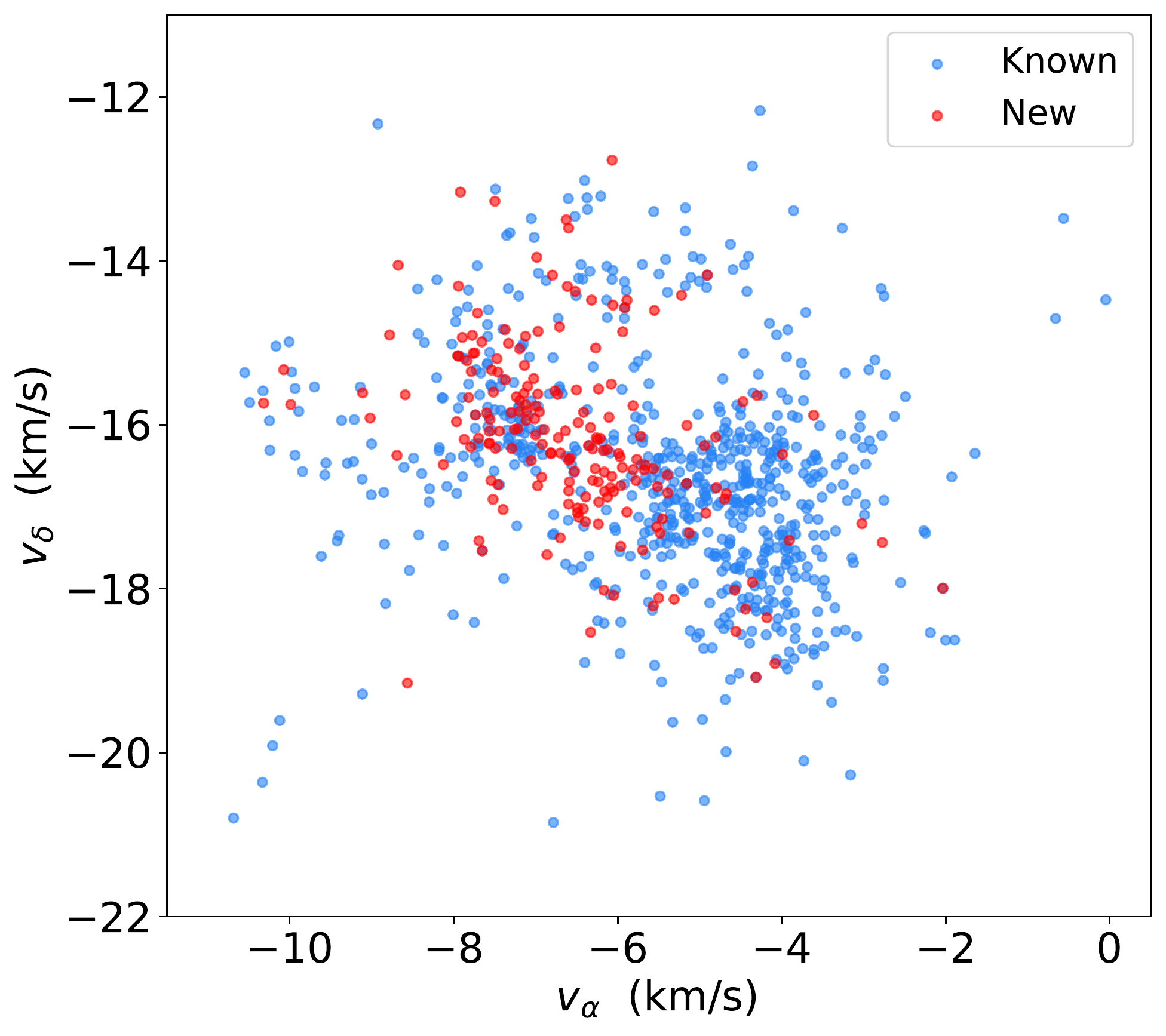}
    \caption{Tangential velocities of the $\rho$~Oph sources. The known sources are shown in blue, and the new sources are shown in red.}
    \label{fig:pmradec}
\end{figure}

Figure~\ref{fig:pmradec} shows the tangential velocity distribution of the $\rho$~Oph sources. The impostor sources (see Sect.~\ref{sec:imp}) from the literature are not included in the diagram, to avoid the influence of outliers. Although the distribution of the new sources shows an overlap with the bulk of the known sources around $-6 < v_{\alpha}<-3$ and $-19 < v_{\delta}<-16$, a large part of the population is shifted toward more negative values of $v_{\alpha}$ and less negative $v_{\delta}$, hinting at more than a single population. These two separate dynamical populations can already be recognized in the known sources alone, while the new sources further add to the second dynamical subgroup around $-10 < v_{\alpha}<-6$ and $-14 < v_{\delta}<-18$.

\begin{figure*}[!t]
    \centering
    \includegraphics[width=\hsize]{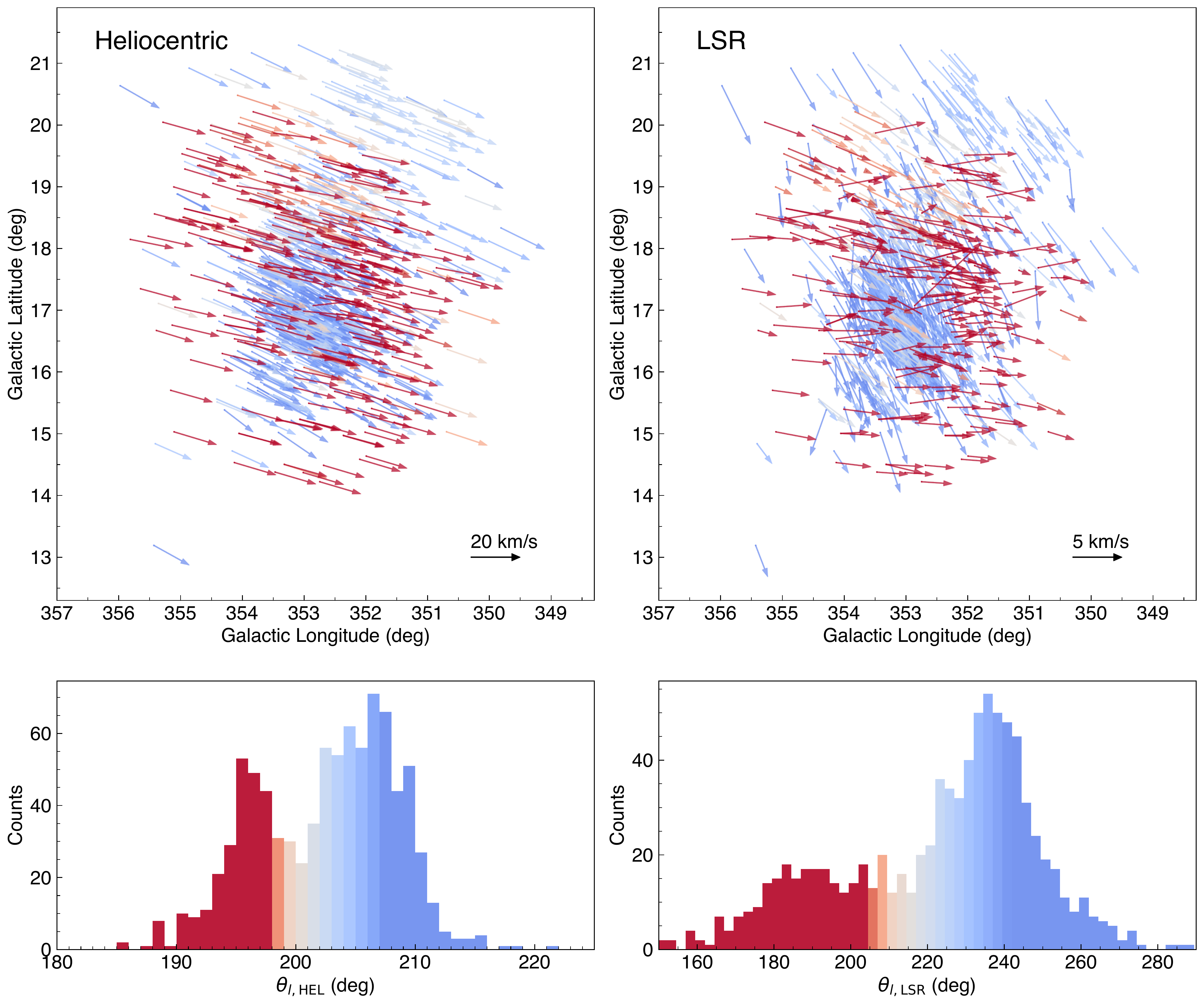}
    \caption{Analysis of the two populations in $\rho$ Oph based on their proper motion. 
    \textit{Top row:} Galactic distribution of the known and new $\rho$\,Oph members, including all new sources (without stability cut), while impostors are excluded. Arrows represent the tangential velocity vectors, color-coded for the angle between the vectors and the $l$-axis ($\theta_{l,\mathrm{HEL}}$ and $\theta_{l,\mathrm{LSR}}$ in left and right panel, respectively). The left panel shows the heliocentric tangential velocity vectors ($\vec{v}_\text{HEL}$), as derived from \textit{Gaia} EDR3 parameters ($v_l$, $v_b$), the right panel shows the tangential velocity vectors relative to the local standard of rest ($\vec{v}_\mathrm{LSR}$ based on $v_{l,\mathrm{LSR}}$, $v_{b,\text{LSR}}$).
    The black arrows in the bottom right corners indicate the vector length for velocities of 20\,km/s and 5\,km/s for $\vec{v}_\text{HEL}$ and $\vec{v}_\text{LSR}$, respectively. These reference vectors have an angle of 180$^\circ$ relative to the $l$-axis.
    \textit{Bottom row:} Histograms showing the distributions of angles $\theta_{l,\mathrm{HEL}}$ and $\theta_{l,\mathrm{LSR}}$ for the sources as in the top panels. The bins in the left histogram have a width of $1^\circ$ and in the right histogram of $2.5^\circ$ since $\theta_{l,\mathrm{LSR}}$ covers a larger range of angles.   
    The histograms are color-coded for the angles as in the top panels.}
    \label{fig:ppmangle_arrows}
\end{figure*}

For further analysis of this distinct kinematic subgroup, we determined the proper motions in Galactic coordinates and the angles between the Galactic proper motion vectors ($\vec{\mu}_{l,b}$) and the $l$-axis ($\theta_{l,\mathrm{HEL}}$) in the heliocentric reference frame, and added these values to our table in a new column for all the sources with proper motion measurements. Analyzing these angles in a histogram reveals the two dynamically different populations as two distinct peaks, as can be seen in the histogram in Fig.~\ref{fig:ppmangle_arrows} in the bottom left image. To disentangle these two populations, we use the angle distribution as a visual aid and apply a cut of $\theta_{l,\mathrm{HEL}} < 200^\circ$, resulting in a subgroup of 304 sources for the second population when excluding 2 impostor sources. Using the proper motion angles relative to the local standard of rest ($\theta_{l,\mathrm{LSR}}$) produces a similar result, as shown in the bottom right image of Fig.~\ref{fig:ppmangle_arrows}. However, using this method, the separation between the two populations is not as evident, indicating that there might be more than two dynamical populations. For simplicity, we considered only two populations in our work and refer to future studies on Sco-Cen (Ratzenb\"ock et al., in prep.) for a more detailed analysis.

Figure~\ref{fig:ppmangle_arrows} highlights the influence of the Sun's reflex motion on the heliocentric proper motions. The top panels show the direction of motion using the heliocentric velocities (left) and the direction of motion when correcting for the Sun's motion (right), showing velocities relative to the local standard of rest (LSR). The latter show a less clear separation between the two populations. To separate the two populations, we used the heliocentric proper motion to avoid injecting in the final selection uncertainties related to the Sun's motion \citep{Schoenrich2010}. In any case, making a selection of the populations in $\theta_{l,\mathrm{LSR}}$ would not change the result significantly.

For further discussion, this second dynamically distinct population shall be referred to as ``Pop~2,'' while the remaining shall be referred as ``Pop~1'' sources, after excluding impostors (see Sect.~\ref{sec:imp}). We define the sources in Pop~1 to be all sources from our $\rho$~Oph catalog excluding impostors and Pop~2 sources. This population comprises the clusters of young stars around the $\rho$~Ophiuchi star and the main Ophiuchus clouds (L1688, L1689, L1709). Concluding, we identify 304 sources in Pop~2 and 1022 in Pop~1 when including sources of all stabilities. When applying a cut at stability > 4\% for the new sources, we are left with 296 sources in Pop~2 and 993 sources in Pop~1 (see Table~\ref{tab:overview-numbers}).

The 304 sources in Pop~2 coincide with the sources whose tangential velocities create the second dynamical structure in Figure~\ref{fig:pmradec}. In other words, the two subpopulations seen in this figure and the bimodal angle distribution consist of the same stars. 
115 of these 304 sources (37.8\,\%) are new sources identified by the algorithm. Further examination of this subgroup reveals that unlike Pop~1, Pop~2 sources are mostly dispersed and are distributed relatively evenly all around the core of the cloud (see Fig.~\ref{fig:map}). Their distances exhibit a similar distribution to the other $\rho$~Oph sources, which shows that the two populations occupy approximately the same 3D volume. 

\begin{table}[t!]
    \centering
    \caption{Average astrometric properties, including their standard deviations (1$\sigma$), for the two populations (Pop~1 and Pop~2) in the $\rho$~Oph region.}
    \begin{tabular}{lrr}
    \hline
    \hline
    Dimension & Pop~1 & Pop~2 \\
    \hline
$\alpha$ (deg) & 246.4$\pm$1.3 & 246.0$\pm$1.2 \\
$\delta$ (deg) & -24.2$\pm$0.8 & -23.9$\pm$1.4 \\
$\varpi$ (mas) & 7.1$\pm$0.4 & 7.1$\pm$0.4 \\
$d$ (pc) & 140.4$\pm$8.0 & 141.3$\pm$7.9 \\
$\mu_{\alpha}^{*}$ (mas/yr) & -7.2$\pm$2.1 & -11.4$\pm$1.9 \\
$\mu_{\delta}$ (mas/yr) & -25.3$\pm$2.3 & -23.4$\pm$2.1 \\
$v_{\alpha}$ (km/s) & -4.7$\pm$1.1 & -7.6$\pm$1.2 \\
$v_{\delta}$ (km/s) & -17.0$\pm$1.4 & -15.7$\pm$1.2 \\
$v_r$ (km/s) & -6.2$\pm$4.5 & -3.9$\pm$3.3 \\
$X$ (pc) & 132.8$\pm$7.7 & 133.5$\pm$7.5 \\
$Y$ (pc) & -16.2$\pm$2.0 & -16.3$\pm$2.9 \\
$Z$ (pc) & 42.3$\pm$3.8 & 42.8$\pm$3.9 \\
$U$ (km/s) & -5.5$\pm$3.4 & -4.1$\pm$3.0 \\
$V$ (km/s) & -15.1$\pm$1.3 & -16.2$\pm$1.5 \\
$W$ (km/s) & -9.4$\pm$1.4 & -5.7$\pm$1.5 \\

    \hline
    \end{tabular}
    \label{tab:avgpop2}
\end{table}

Table~\ref{tab:avgpop2} shows the average values of the distances, proper motions, radial velocities, Galactic Cartesian positions $X,Y,Z$ and Galactic Cartesian velocities $U,V,W$, and the standard deviations of these parameters for the two populations (Pop~1 and Pop~2) in the $\rho$~Oph region. The average 3D positions of the two populations only exhibit small deviations, showing that they are not merely a 2D overlap, but mixed in all three spatial dimensions. As can be seen from the proper motions and tangential velocities in Table~\ref{tab:avgpop2}, the Pop~2 sources exhibit slightly different dynamical properties, which set them apart. Although the $U$ and $V$ velocities of the two populations hardly differ from each other, they occupy different regions in the $UVW$ velocity space because of the larger differences in $W$. The bimodality seen in Fig.~\ref{fig:pmradec} can also be seen in the $UVW$ space; however, only 55 sources (18.1\,\%) from the second population have $UVW$ velocities. By computing the difference between the $UVW$ vectors of the two populations, we find that they are moving away from each other at about 4.1 km/s and will no longer overlap in about 4 Myr.


\begin{figure*}[p!]
\begin{minipage}[t]{1\linewidth}
\centering
\includegraphics[width=\textwidth]{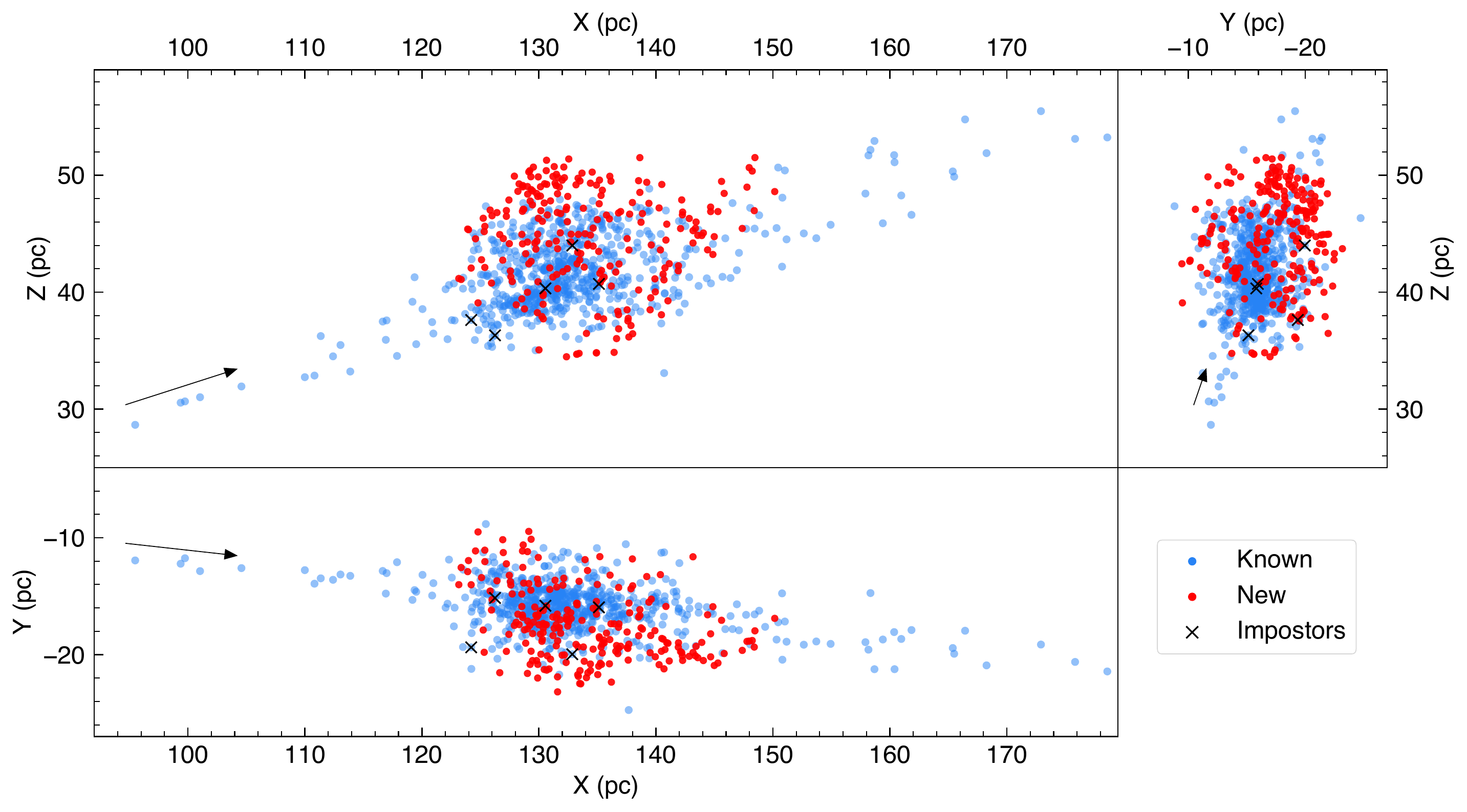}
\caption{Heliocentric Galactic Cartesian coordinates of the $\rho$~Oph sources. The known sources from the literature are marked with blue dots, the new sources with red dots, and impostor sources (Sect.~\ref{sec:imp}) with black crosses (see legend). No quality or stability criteria were applied to the displayed sources. The black arrows in each panel indicate the line-of-sight from the Sun, pointing toward the star $\rho$~Oph and plotted from $d=100$--110\,pc, which results in different arrow lengths due to projection effects. An interactive 3D version is available online at \url{https://homepage.univie.ac.at/josefa.elisabeth.grossschedl/rhoOph-Fig6.html}.}
\label{fig:xyz}
\end{minipage}
\vspace{1cm}
\begin{minipage}[b]{1\linewidth}
\centering
\includegraphics[width=\textwidth]{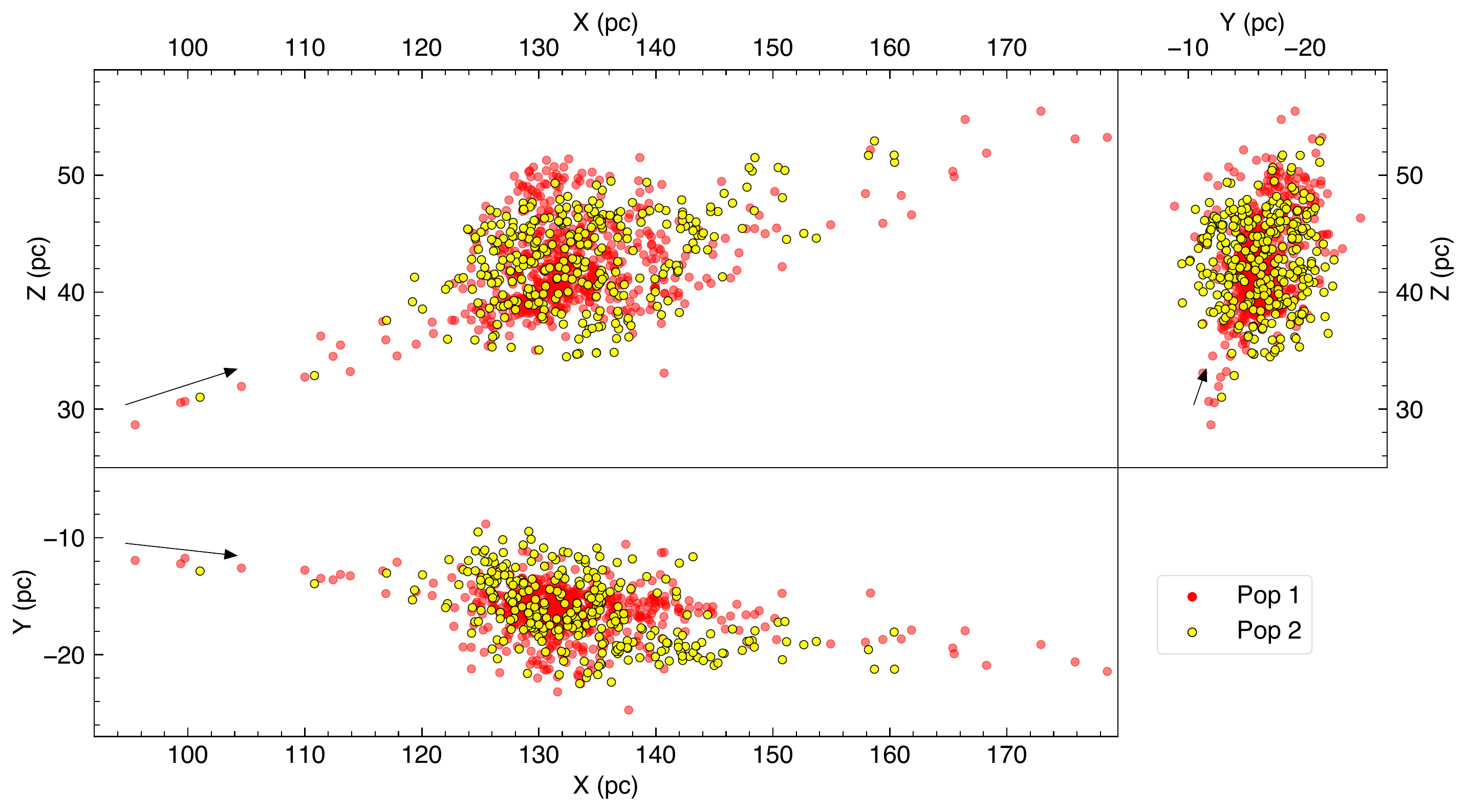}
\caption{Heliocentric Galactic Cartesian coordinates of the $\rho$~Oph sources, separated into Pop~1 (red) and Pop~2 (yellow). No quality or stability criteria were applied to the displayed sources. As in Fig.~\ref{fig:xyz}, the black arrows in each panel indicate the line-of-sight from the Sun. An interactive 3D version is available online at \url{https://homepage.univie.ac.at/josefa.elisabeth.grossschedl/rhoOph-Fig7.html}.}
\label{fig:xyz_pop12}
\end{minipage}
\end{figure*}


Figure~\ref{fig:xyz} shows the Galactic Cartesian coordinates of the known and new $\rho$~Oph sources for a visualization of their 3D distribution. The literature sources exhibit a more elongated distribution.
Previously labeled impostors in Sect.~\ref{sec:imp} are marked with black crosses in Fig.~\ref{fig:xyz}. The elongation is most prominent along the line-of-sight, which is mostly caused by the larger errors in the parallax measurements compared to celestial coordinates, while some of the elongation could be caused by outliers. It can be seen in Fig.~\ref{fig:xyz} that the new sources are rather distributed at the outskirts of the main cluster, indicating that they have been missed previously because they are more dispersed in space. 

In Figure~\ref{fig:xyz_pop12} we show the same Galactic Cartesian representation as in Fig.~\ref{fig:xyz}, this time highlighting the 3D distribution of the Pop~1 and Pop~2 sources. It can be seen that the two populations largely occupy the same space, while there is a lack of Pop~2 sources at very high $Z$ when compared to Pop~1, best visible in the $X$ versus $Z$ panel. This distribution is consistent with the projected Galactic distribution in Fig.~\ref{fig:map}, where a similar lack of Pop~2 sources to the Galactic north can be seen.

\subsection{Observational HRD} \label{sec:cmd}

\begin{figure*}[ht!]
\begin{minipage}[b]{0.5\linewidth}
\centering
\includegraphics[width=\textwidth]{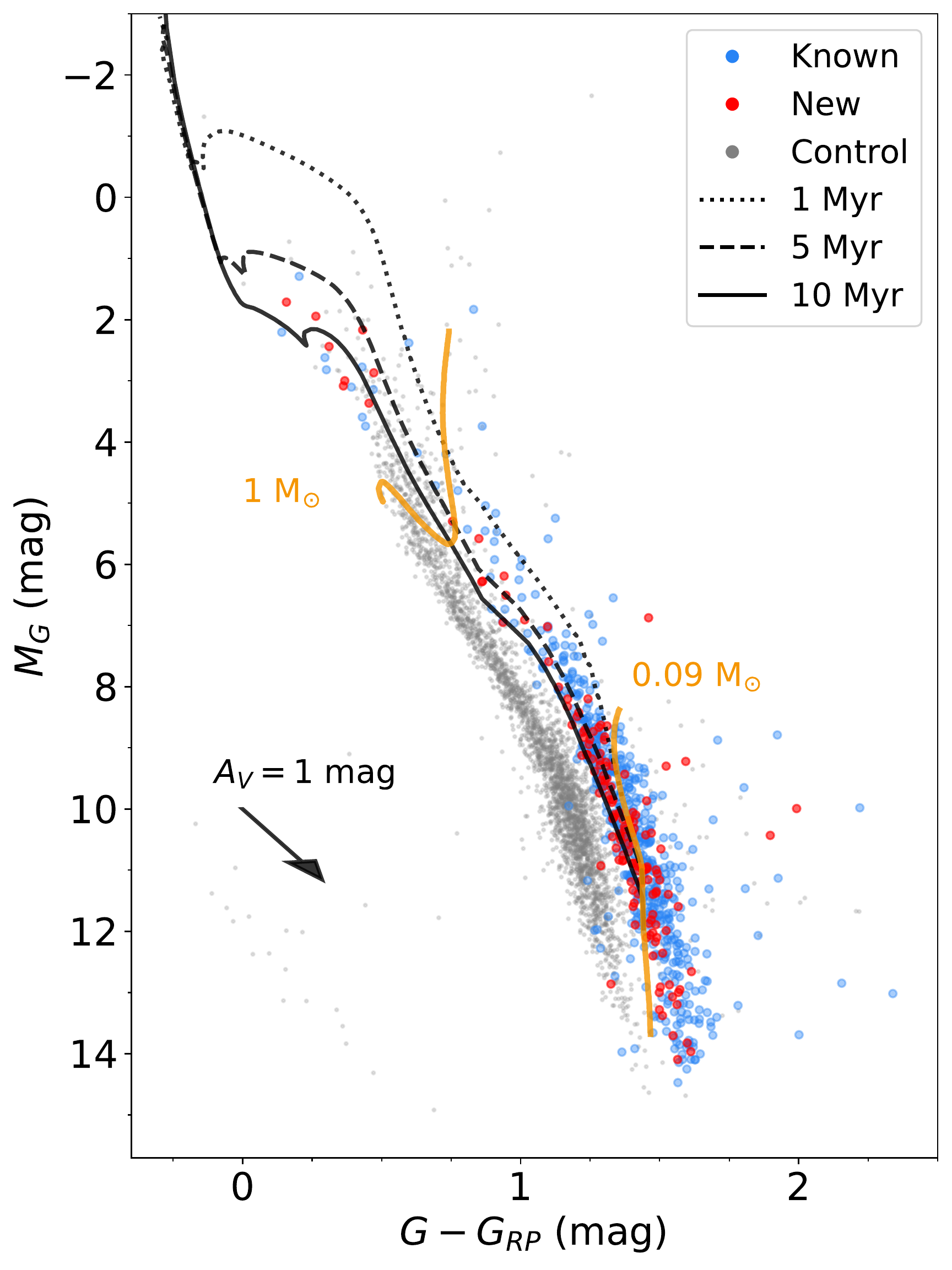}
\end{minipage}
\hspace{0.0cm}
\begin{minipage}[b]{0.5\linewidth}
\centering
\includegraphics[width=\textwidth]{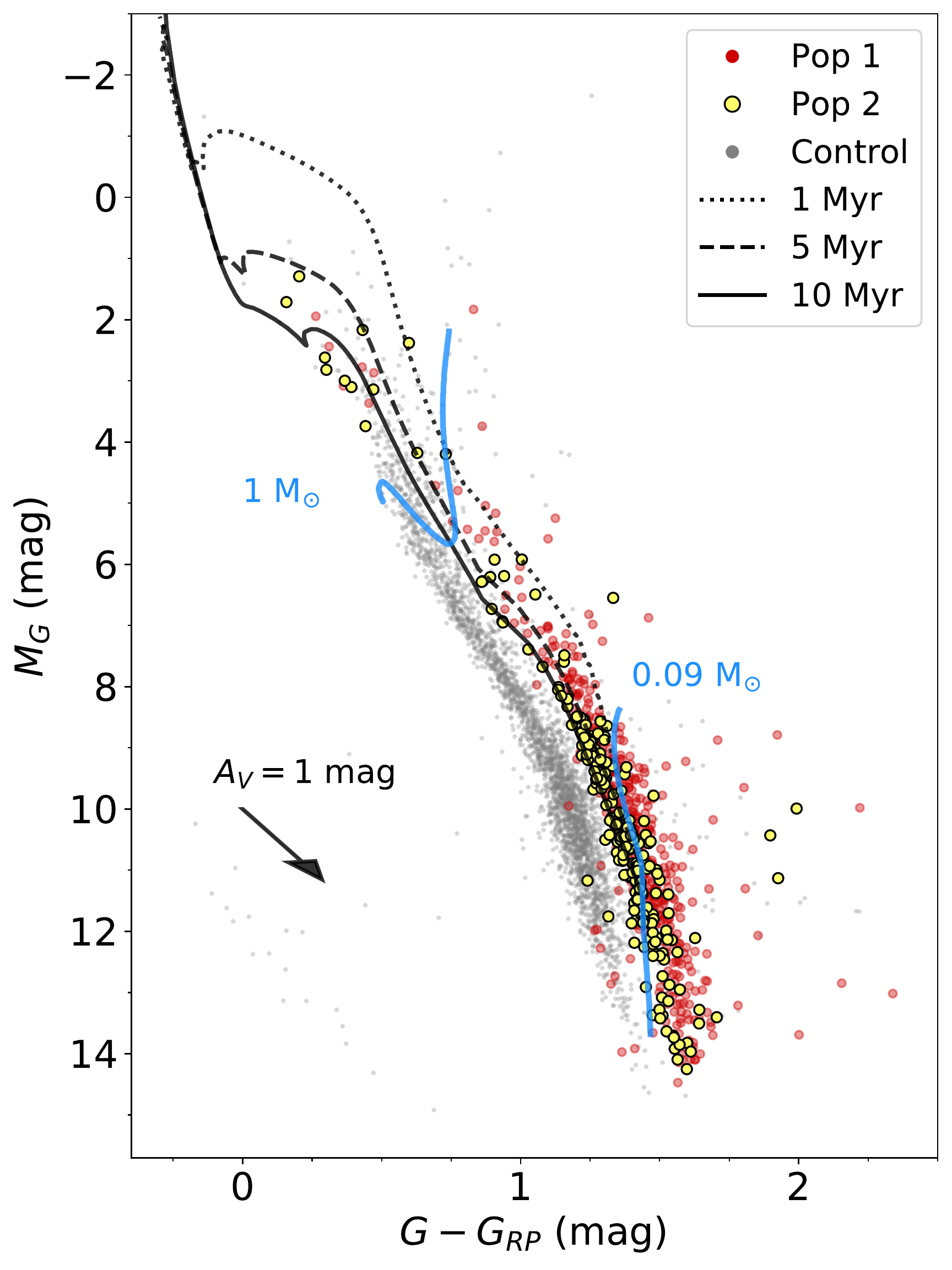}
\end{minipage}
\caption{%
Observational HRDs using the \textit{Gaia} $G$ and $G_{RP}$ passbands, with corrections applied to the $G$ passband. An extinction vector with $A_\mathrm{V} = 1\,$mag is shown as black arrow. The isochrones correspond to ages of 1, 5 and 10\,Myr (see legend). The iso-mass lines (left: orange, right: blue) for 0.09\,$\text{M}_\odot$ and 1\,$\text{M}_\odot$ include stars with ages from 0.1 to 100\,Myr. A control sample is shown in gray in the back. 
\textit{Left:} Comparing known (blue) and new (red) sources in $\rho$~Oph. 
\textit{Right:} Comparing the two populations in $\rho$~Oph, with Pop~1 in red and Pop~2 in yellow.}
\label{fig:cmd}
\end{figure*}

Figure \ref{fig:cmd} (left) shows an observational Hertzsprung-Russel Diagram (HRD) of the $\rho$~Oph sources with the known sources in blue and the new ones in red. To create the diagram we use the \textit{Gaia} EDR3 passbands $G$ and $G_{RP}$, for both the $\rho$~Oph and the control sample. Since \textit{Gaia} EDR3 photometry is affected by systematic errors, corrections were applied to the $G$ band as described in \cite{GaiaEDR32020photocor}. Using the observed magnitudes $m_G$ in the $G$ band and the individual distances $d$ of the sources, we computed the absolute magnitudes $M_G$ in the $G$ band with $M_G=m_G+5-5\log_{10}d$. Quality cuts as described in Appendix~\ref{app:Gaia} were applied to the \textit{Gaia} data of the $\rho$~Oph and control sample to include only high quality photometry and astrometry.
Isochrones from the PARSEC models \citep{Marigo} for \textit{Gaia} EDR3 photometry are over-plotted in Fig.~\ref{fig:cmd} for 1, 5, and 10\,Myr. An extinction vector in the $V$ passband, labeled as $A_V$, is shown to visualize the direction and magnitude of extinction in this color-magnitude space using the reddening law from \cite{Cardelli1989} and \cite{ODonnell1994} provided by PARSEC. Two equal-mass-curves for sources with 0.09\,$\text{M}_\odot$ and 1\,$\text{M}_\odot$ are over-plotted.

The distribution of the known and new sources in the left panel of Fig.~\ref{fig:cmd} overlap, indicating similar ages and luminosities, as also described in Sect.~\ref{sec:ophclass} and shown in Fig.~\ref{fig:stab_isochrone}. This further confirms that they belong to the same region. Their distribution is consistent with earlier work of \cite{LuhmanRieke1999} and \cite{Esplin}, who find ages of 0.3--6\,Myr for $\rho$~Oph sources. Most of the new sources are low-mass stars, similar to the known sources, probably consisting mainly of M-type spectral classes or substellar objects. 

In the right panel of Fig.~\ref{fig:cmd} we show a similar observational HRD as in the left panel, 
showing the two dynamical populations in the $\rho$~Oph region. The first population (Pop~1), which comprises the clusters of young stars around the $\rho$~Ophiuchi star and the main Ophiuchus clouds (L1688, L1689, L1709), is shown in red, and the second dynamically distinct population (Pop~2) is shown in yellow. 
One can see that the second population appears to be slightly older than the first and aligns better with older isochrones. To determine the approximate age of the second population, we compute a least mean square fit to the data, as similarly done in Sect.~\ref{sec:ophclass}, using the $G$, $BP$ and $RP$ passbands, to isochrones with solar metallicity from the PARSEC models \citep{Bressan2012}. We use only high-fidelity sources with stability > 4, and quality cuts of ruwe < 1.4 and astrometric\_sigma5d\_max < 0.5 (for definitions of used \textit{Gaia} parameters, see Table~\ref{tab:ophtab}). With this we obtain an approximate age of 10\,Myr for the second population, which is older than the average age of about 5\,Myr of the whole sample.


\subsection{Analysis of infrared colors: Infrared-excess sources}

The evolutionary stages of young stars can be estimated by using IR measurements, which reveal the presence of protoplanetary disks and envelopes around the pre-main-sequence stars. Disks and envelopes emit light in IR wavelengths due to their warm dust emission. Cross-matching our complete $\rho$~Oph catalog with data from \textit{WISE} \citep{WISE}, in our case the \textit{AllWISE} catalog, provides stars with the required IR photometry to analyze IR excesses. We note that not all sources are represented by \textit{WISE}. The cross-match yielded 1110 sources with \textit{WISE} data, which is 82.7\,\% of our $\rho$~Oph sources. The $W1$, $W2$, and $W3$ passbands correspond to wavelengths of $\SI{3.4}{\micro\meter}$, $\SI{4.6}{\micro\meter}$, and $\SI{13}{\micro\meter}$, respectively. To use only high quality measurements in our diagram, we only included sources above a specific signal-to-noise ratio (S/N). Sources had to fulfill $w1snr>10$, $w2snr>10$, and $w3snr>7$ for the $W1$, $W2$, and $W3$ passbands. This cut was applied to the $\rho$~Oph and the control sample, leaving 750 sources for the diagram, which is 55.8\,\% of the total  $\rho$~Oph sample.

\begin{figure*}[ht]
\begin{minipage}[b]{0.48\linewidth}
    \centering
    \includegraphics[width=\columnwidth]{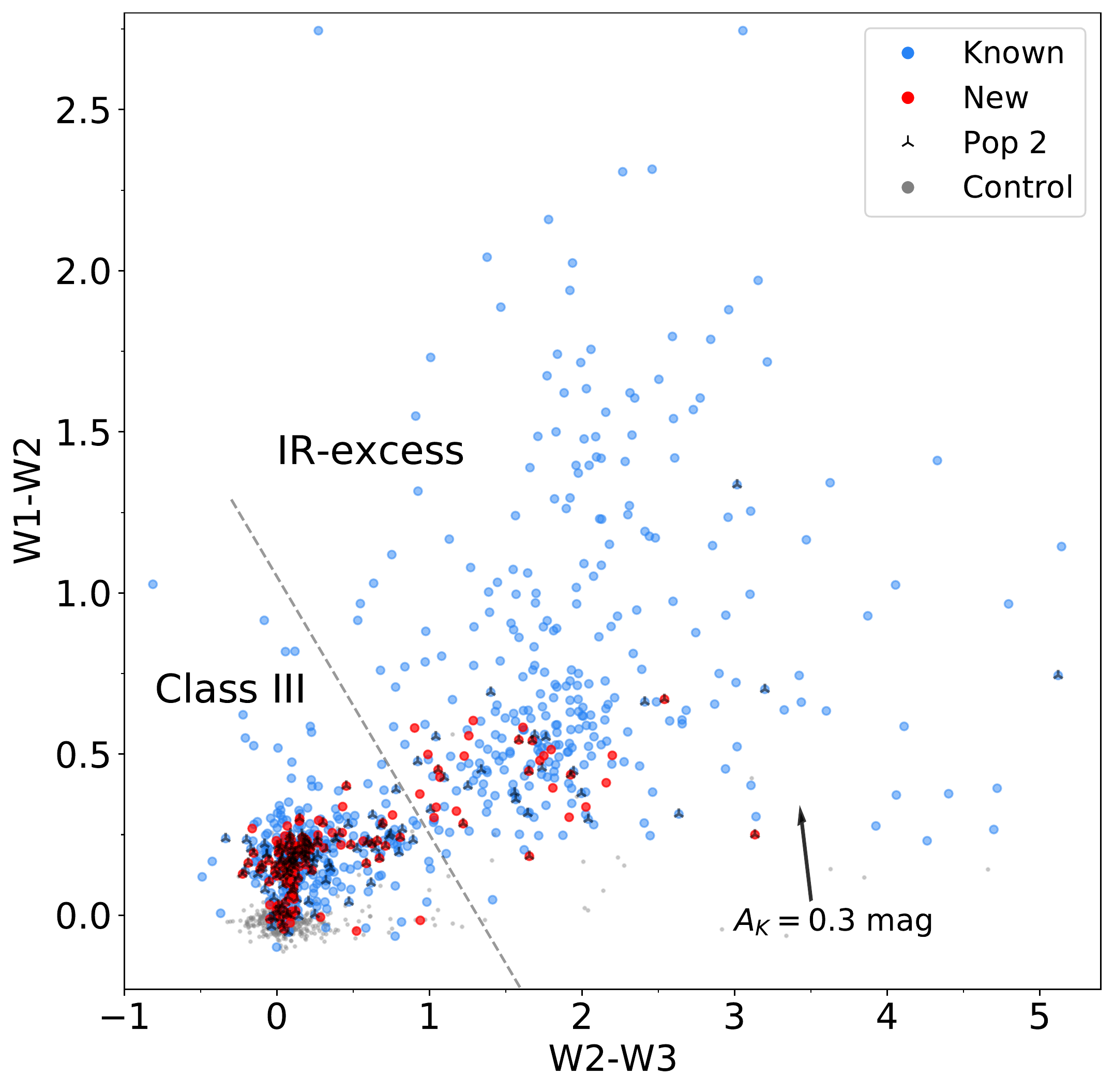}
    \caption{Mid-infrared color-color diagram of the known $\rho$~Oph sources in blue and the new ones in red, including the control sample in gray, using the $W1$, $W2$, and $W3$ passbands from the \textit{WISE} catalog. The sources comprising the second population (Pop~2) are marked by black symbols. An extinction vector in the $K_\mathrm{S}$ passband, labeled as $A_\mathrm{K}$, is also included. The sources above the dashed line with $W1-W2 > 1.05 -0.8 \cdot (W2-W3)$, are YSOs with IR excess due to a circumstellar disk (Class\,I or Class\,II), while those below the line are Class\,III YSOs.}
    \label{fig:w123}
\end{minipage}
\hspace{0.5cm}
\begin{minipage}[b]{0.48\linewidth}
    \centering
    \includegraphics[width=\columnwidth]{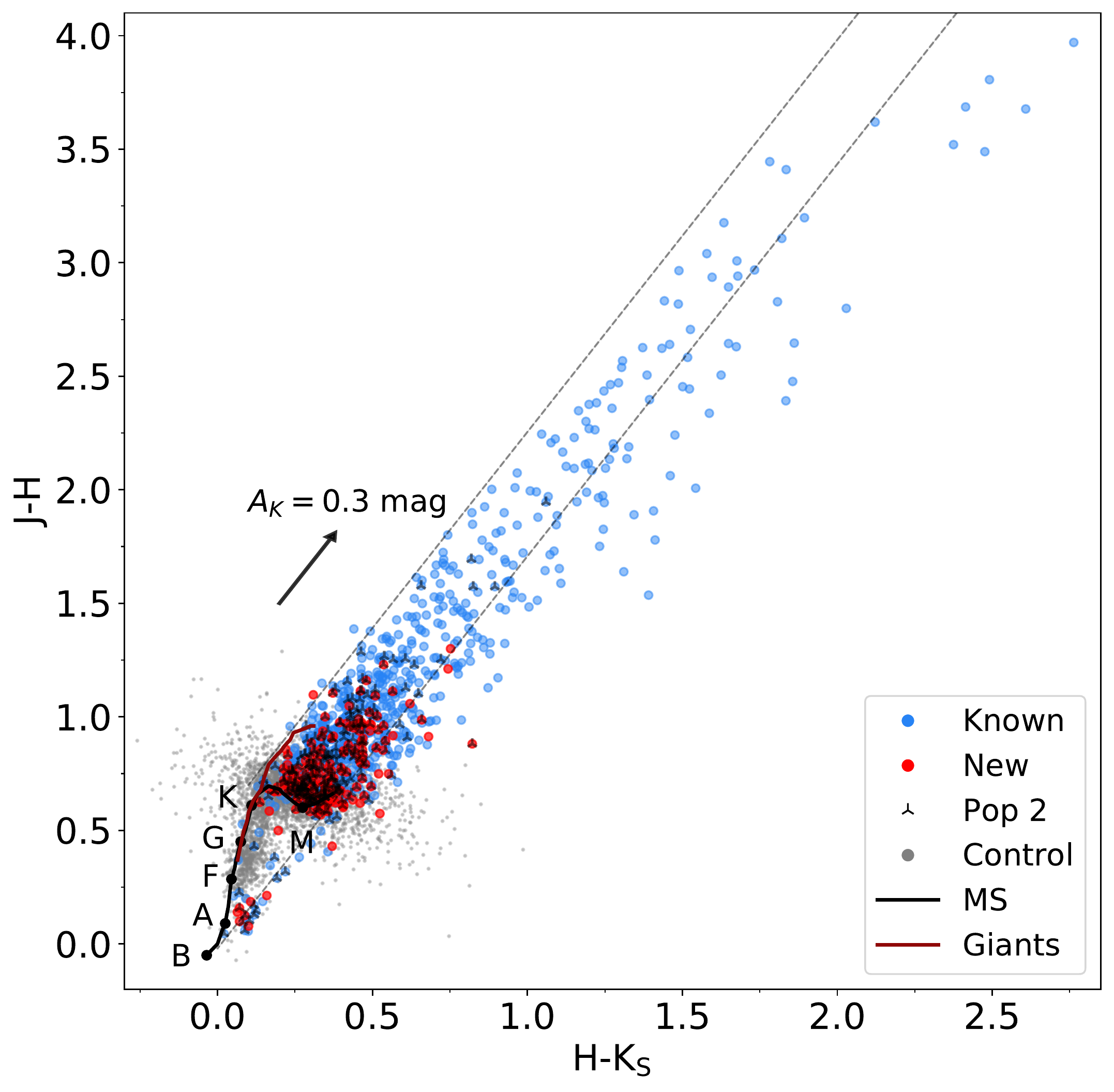}
    \caption{Near-infrared color-color diagram of the know (blue) and new (red) $\rho$~Oph sources, including the control sample in gray and the sources from the second population (Pop~2) as black symbols, using the $J$, $H$ and $K_\mathrm{S}$ passbands from 2MASS. The main sequence and the giant branches from \cite{Bessell1988} are included in the diagram, as well as an extinction vector in the $K_\mathrm{S}$ passband, labeled as $A_\mathrm{K}$. The two parallel lines with the slope of the extinction vector enclose sources that are reddened mainly due to extinction in this color space.}
    \label{fig:jhk}
\end{minipage}
\end{figure*}

Figure~\ref{fig:w123} shows a color-color diagram for $W1-W2$ versus $W2-W3$, with the known sources in blue and the new ones in red. The control sample is included in gray, and the sources of the second population (Pop~2) are marked by black symbols. The extinction vector in the $K_\mathrm{S}$ passband, labeled as $A_\mathrm{K}$, was determined using the reddening law for the $W1$, $W2$, and $W3$ passbands as in \cite{Meingast2018}. A dashed line, serving as a rough estimate, separates two regions in the diagram, namely those with and without IR excess, as similarly done in \cite{Koenig2014}. The functional form of the dashed line is given by $W1-W2 = 1.05 -0.8 \cdot (W2-W3)$. Sources further to the top and right in the diagram exhibit an IR excess and are therefore most likely YSOs with envelopes or circumstellar disks, Class\,I or Class\,II, while Class\,II are similar to Classical T Tauri stars \citep{Greene}.

Most of the new sources have little or no IR excess, which could be the reason why they have not been identified in any previous IR surveys. Sources below and to the left of the dashed line in Fig.~\ref{fig:w123} are either Class\,III YSOs or main sequence stars. As Figure~\ref{fig:cmd} confirms that $\rho$~Oph consists mainly of young stars, this implies that the $\rho$~Oph sources below the line can only be Class\,III YSOs, which are associated with tenuous disks or bare photospheres, therefore creating no detectable infrared excess \citep{Canovas}. 

As can be seen from the red sources above and to the right of the dashed line in Fig.~\ref{fig:w123}, we have found 28 new sources with IR excess, which are likely Class\,II candidates. This corresponds to a disk fraction of about 19.9\,\% in the new sources, considering the displayed 141 new sources in the diagram.
The known sources contain both Class\,I and Class\,II candidates. The fraction of sources with IR excess in the known population is roughly 48.6\,\%, considering the 609 known sources within our \textit{WISE} quality criteria, with 313 Class\,III YSOs and 296 YSOs with IR excess. Further analysis of the 28 new YSOs with IR excess reveals that they are located further away from the core of the cloud, which might explain why they have not been found in any previous IR study of $\rho$~Oph, which focused mainly on the core region. 19 of the 28 new disk sources are from Pop~1, while 9 of them belong to Pop~2. The positions of the new IR excess sources in the HRD align well with most of the other new sources, showing very little scatter.

The distribution of Pop~2 members in Fig.~\ref{fig:w123} shows deviations from the average, with only 30 YSOs with IR excess and 154 Class\,III sources in the diagram, corresponding to a disk fraction of 16.3\,\%, while Pop~1 has 293 YSOs with IR excess and 260 Class\,III sources in the diagram, resulting in a larger fraction of sources with IR excess of 53.0\,\%. We conclude that Pop\,2 contains overall more evolved stellar members and is likely at a later evolutionary stage compared to Pop~1 since the majority do not show any IR excess. This is consistent with the older age of Pop~2 seen in the optical HRD (Fig.~\ref{fig:cmd}, right panel). We note that the fraction of sources with IR excess could be overestimated for Pop~1 since even sources without proper motion values were counted to Pop~1, as defined in Sect.~\ref{subsec:ast}. Therefore, sources without measured astrometry are highly uncertain Pop~1 members since some could belong to Pop~2 or could even be galaxies, which could contaminate an IR-selected YSO sample. 

We note that there are two known sources from Pop~2 that show untypically red colors compared to most other Pop~2 sources. The source with the largest $W1-W2$ value has a \textit{Gaia} source ID of \verb|6049129800518036992|, and it is located near the core of the molecular cloud. Based on its color, it could be a flat-spectrum source or Class\,I (protostar). The proper motion direction indeed seems to fit to the Pop~2 sample; however, after checking the source in more detail, we find that the source has overall larger errors, indicating that its proper motion and distance, hence the tangential velocity, could be dominated by errors. Therefore, the Pop~2 membership of this source is uncertain, and it could be part of the younger Pop~1. This would reduce the disk percentage of Pop~2 down to 14.8\,\%. The other Pop~2 source with a very significant IR excess, namely the one with the largest $W2-W3$ value at the right of the diagram, has a \textit{Gaia} source ID of \verb|6050279163829546112|. The IR excess in W3 could indicate that the source is a transition disk.

We conclude that we have found 28 new YSOs with IR excess and 113 new Class\,III YSOs in Ophiuchus. The fraction of IR excess sources to Class III YSOs is around 0.25 in the new sources, 0.95 in the known sources, and around 0.76 in the entire population. Again, the fraction of known sources with IR excess could be slightly overestimated due to above mentioned reasons. An overview of the final numbers is given in Table~\ref{tab:overview-numbers}. All sources with IR excess (Class\,I or Class\,II) according to Fig.~\ref{fig:w123}, in total 324, are marked in our final catalog in the column ``IR\_excess'' with a ``1,'' while the remaining sources (Class\,III) are marked with a ``0.'' Sources not included in Fig.~\ref{fig:w123} are not classified in this work.

Cross-matching our complete $\rho$~Oph sample with data from 2MASS \citep{2MASS} provides us with further IR measurements in the $J$, $H$, and $K_\mathrm{S}$ passbands, which correspond to wavelengths of $\SI{1.25}{\micro\meter}$, $\SI{1.65}{\micro\meter}$, and $\SI{2.17}{\micro\meter}$, respectively. Figure~\ref{fig:jhk} shows a color-color diagram of $H-K_\mathrm{S}$ versus $J-H$. In order to show only high quality measurements, we use the quality cuts $j\_cmsig, h\_cmsig, k\_cmsig < 0.1$. The known, new, and control sources are in blue, red, and gray, respectively, while the sources from the second population (Pop~2) are marked by black symbols. The main sequence (MS) and giant branches are included in the diagram, as determined by \cite{Bessell1988}. The extinction vector in the $K_\mathrm{S}$ passband, labeled as $A_\mathrm{K}$, was determined using the reddening law for the $J$, $H$ and $K_\mathrm{S}$ passbands by \cite{Meingast2018}. Two parallel lines with the slope of the extinction vector were added to enclose reddened sources above the main sequence.

As can be seen from their positions in Figure~\ref{fig:jhk}, most of the known and new sources are M stars. However, there are also several higher-mass stars among the new sources, as seen in the bottom left of the diagram. We find that these stars are located relatively far from the core of the cloud, which could explain why they have not been added as members in previous studies. Furthermore, the new sources are, on average, less extincted than the known ones, as would be expected, since they were selected based on the \textit{Gaia} catalog.

\begin{figure*}[t!]
    \centering
    \includegraphics[width=\hsize]{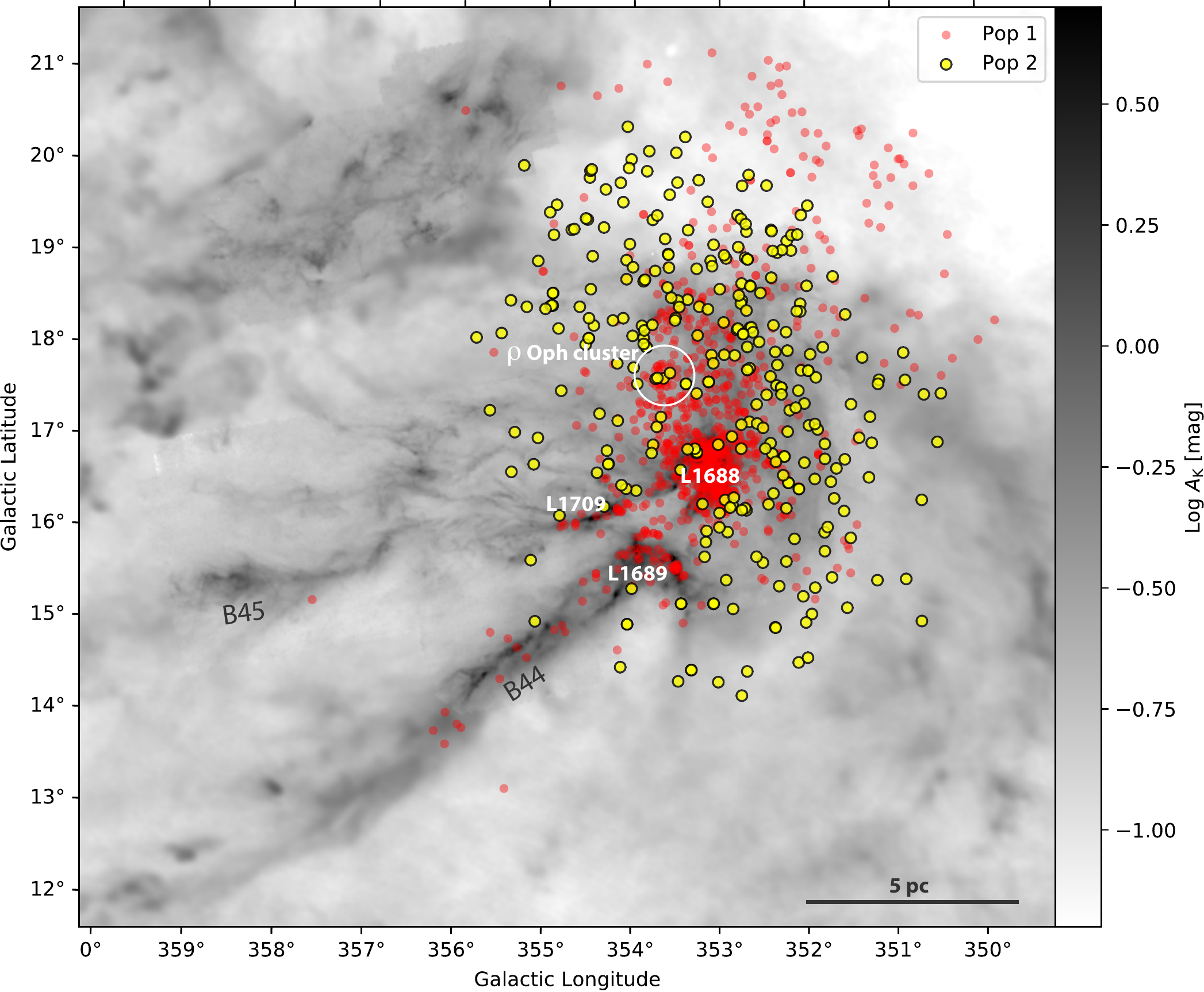}
    \caption{Spatial distribution of the two dynamical populations in $\rho$~Oph in red and yellow circles. The $\rho$~Oph cluster, centered on the $\rho$~Ophiuchi star, is marked by a white open circle. The actively star-forming clouds, L1688, L1689, and L1709, are also marked. Impostors (see Sect.~\ref{sec:imp}) are not included in this figure, whereas low stability sources are. The background grayscale is a column density map of Ophiuchus made with \textit{Herschel}, \textit{Planck}, and 2MASS data (Alves et al., in prep).}
    \label{fig:map}
\end{figure*}

\begin{table*}[t!]
    \caption{Final numbers of sources resulting from our $\rho$\,Oph stellar member analysis.}
    \centering
    \begin{tabular}{lr}
    \hline
    \hline
    \multicolumn{1}{c}{(Sub)sample} &
    \multicolumn{1}{c}{N} \\
    \hline
Known literature selected sources & 1114 \\
Literature selected sources with measured \textit{Gaia} EDR3 parallax & 682 \\
Literature selected sources without impostors & 1097 \\
Impostor sources in the literature & 17 \\
\hline
All new sources without stability cut & 229 \\
New sources with stability cut & 191 \\
New sources with circumstellar disks (Class\,II) & 28 \\
New Class\,III sources & 113 \\
\hline
Total number of $\rho$\,Oph sources without stability cut & 1343 \\
Total number of $\rho$\,Oph sources with stability cut for new sources & 1305 \\
Total number of $\rho$\,Oph sources without impostors & 1326 \\
Total number of $\rho$\,Oph sources with stability cut for new sources and without impostors & 1288 \\
\hline
Pop~1 sources without stability cut & 1022 \\
Pop~1 sources with stability cut for new sources & 993 \\
Pop~2 sources without stability cut & 304 \\
Pop~2 sources with stability cut for new sources & 296 \\
\hline

    \end{tabular}
    \label{tab:overview-numbers}
\end{table*}

\section{Discussion} \label{sec:discussion}

In this work we applied the classification strategy developed by \cite{Ratzenboeck} to identify new members of the $\rho$~Oph region in \textit{Gaia} EDR3. This method yielded 191 new high stability members with similar properties in position and motion to the 1114 known sources from the literature. From these results, we were able to create a master catalog of all known sources in $\rho$~Oph, including our new sources from \textit{Gaia} EDR3. This so far most complete sample of $\rho$~Oph contains 1305 sources (or 1343 when also including the new sources with a stability $< 4$).

\subsection{The $\rho$~Oph region is a mixture of two young populations} \label{subsec:2pops}

The tangential velocity distribution of the final sample, presented in Fig.~\ref{fig:pmradec}, reveals structure hinting at the presence of more than one population. The bimodal distribution of the proper motion angles presented in Fig.~\ref{fig:ppmangle_arrows} further asserts the existence of two main populations in the surveyed area, which we call Pop~1 and Pop~2. What is discussed in the literature as the ``$\rho$~Oph star-forming region'' or ``$\rho$~Oph core'' is in fact a mixture of at least two populations, with similar but distinct dynamical properties and ages, occupying approximately the same 3D volume. The first (Pop~1), with ages 0.3--6 Myr \citep{LuhmanRieke1999,Erickson2011-vu,Esplin}, as confirmed in Fig.~\ref{fig:cmd}, comprises clusters of young stars around the $\rho$~Ophiuchi star and the main Ophiuchus clouds, namely L1688, L1689, L1709 (see Fig.~\ref{fig:map}). The second population (Pop~2) appears more dispersed in comparison and has an older age up to $\sim 10\,$Myr, a disk fraction of $\sim 16.3\,\%$, and 3D motions of $U, V, W = -5.5, -16.2, -5.7\,$ km/s. Given that the age, disk fraction, and 3D motion are similar to those of Upper Sco ($U, V, W = -5.1, -16.0, -7.2\,$ km/s, disk fraction $\sim 20\,\%$, age $\sim$10\,Myr, \citealt{PecautMamajek2016,Luhman2020}), it is possible that the 304 Pop~2 sources in the dispersed population originate from the much larger Upper Sco population toward the Galactic north. However, we note that the sources from Pop~2 appear to be cut off toward the Galactic north, as can be seen in Figs.~\ref{fig:xyz_pop12} and \ref{fig:map}. Considering that Upper Sco lies in the north of $\rho$~Oph, it seems unclear if Pop~2 really originates from there. Still, the proper motion of Pop~2 is essentially the same as the proper motion of Upper Sco \citep{Luhman2020}, making it highly unlikely that Pop~2 is not associated with Upper Sco (same age, distance, and motions). More likely, because the training set consists of 77.3\,\% Pop~1 sources, it is possible that this bias caused the algorithm to find fewer Pop~2 sources, causing the apparent cutoff. This will be further examined in future work (Ratzenb\"ock et al. in prep.).

The clear kinematic difference between these two populations, only detectable because of the unprecedented accuracy of \textit{Gaia} EDR3, is the main finding of our study as it sheds light on the genesis of the $\rho$~Oph star-forming region. The proper motion distribution found in Fig.~\ref{fig:ppmangle_arrows}, in combination with RVs, translates into a 3D space motion difference between the two populations of about 4.1 km/s. This relative space motion indicates that the regions are moving away from each other and could imply that the origin of the $\rho$~Oph star-forming region is connected to that of the Upper Sco population. A study of the space motion of the two populations is called for as it will give insights on the origin of the different motions.

The closest active star formation region to Earth, the $\rho$~Oph region, remains a natural laboratory for star formation studies, from core formation and collapse to disk formation and evolution into planets. Our work demonstrates how the unprecedented astrometric precision of \textit{Gaia} is revealing the fine dynamical structure of this nearest star-forming regions.

\subsection{Multiple young populations in star-forming regions}\label{subsec:multi-pops}

Our finding in this paper of a mixed population in $\rho$~Oph is similar to the discovery of the foreground population in front of the Orion Nebula \citep{Alves2012,Bouy2014,Chen2020}. Unfortunately, two of the closest benchmark star formation regions to Earth, the $\rho$~Oph region and the Orion Nebula Cluster, are now known to contain multiple young populations, either in projection or intermingled, which complicates the extraction of star formation observables. These two cases are unlikely the exception. Mixed populations are to be expected, for example, in triggered star formation as a previous generation compresses interstellar gas into a new generation of stars. Characterizing the existence of multiple populations in nearby star formation regions is critical because it directly affects the fundamental star formation observables, such as star formation history, rate, efficiency, and the initial mass function (IMF). Looking forward, multiple populations should be looked for in other nearby star-forming regions, and for at least $\rho$~Oph and the Orion Nebula Cluster, they need to be disentangled for a precise description of the basic star formation observables. 

\subsection{Caveats}\label{subsec:caveats}

Some of the literature sources are located off from the center of the cloud, in particular the ones that seem to trace the B44 filament (L1689, L1712, L1759), away from the center of the distribution and toward the lower Galactic east in Figure~\ref{fig:sky}. These sources might be too far from the cluster center to be considered by the algorithm, since the training set is only located near the center of the distribution (Fig.~\ref{fig:sky}). Still, the sources seen in projection onto B44 are also located at the edge of the proper motion distribution, making them even less likely to be predicted. However, since there are only a handful of sources located so far off, this suggests that the algorithm is not missing a significant number of sources toward the filaments B44 and B45.

\subsection{Comparison with previous work using \textit{Gaia} data}\label{subsec:comparison}

\cite{Canovas} applied several clustering algorithms (\verb|DBSCAN, OPTICS, HDBSCAN|) to identify new sources in the $\rho$~Oph region using the \textit{Gaia} DR2 catalog. We have found sources that were not identified as potential members by \cite{Canovas}, despite also running our search algorithm on the \textit{Gaia} DR2 catalog before the availability of \textit{Gaia} EDR3. Our search in only \textit{Gaia} DR2 yielded around 150 new members, depending on how strictly we set our prior assumptions. Finding so many new YSOs in the same data set suggest that our approach is an effective tool for searching for new members of co-moving stellar structures.

\cite{Esplin} used \textit{Gaia} DR2 data and derived proper motions with multi-epoch data from the \textit{Spitzer} Space Telescope to find 155 new young stars, 102 of these associated with the Ophiuchus clouds and 47 with Upper Sco. Unlike our study, \cite{Esplin} did not use multivariate classification techniques to identify new sources, so we attribute the discovery of the 191 new YSOs over their search to tailored classification techniques as the one described in this paper, which are powerful tools to disentangle stellar populations in the high-precision \textit{Gaia}-era data.  

Concluding, the algorithm from \cite{Ratzenboeck} has shown to be an effective method for identifying stars belonging to a particular population, based on the properties of a subsample of known sources. The method was able to identify 191 new optically visible sources in $\rho$~Oph, providing more information on the optically revealed population of the region. Therefore, we conclude that our method is a useful tool suitable for similar research in the future.

\section{Conclusions}
The main results from this work can be summarized as follows:

\begin{enumerate}
    \item We searched the literature to construct a catalog of 1114 known YSOs toward the $\rho$~Ophiuchi region. We cross-match this catalog with the \textit{Gaia} EDR3, \textit{Gaia}-ESO, and APOGEE-2 surveys and use it to feed a classification algorithm designed to find new, co-moving population candidates in \textit{Gaia} EDR3 using a training set of 150 sources.
    \item We found 191 new YSO candidates in \textit{Gaia} EDR3 belonging to the $\rho$~Ophiuchi region (229 new YSOs including low-fidelity members). The distribution of the new sources in an HR-diagram is very similar to previously known young stars in the region, validating our selection. 
    \item The new sources appear to be mainly Class III M stars and substellar objects, and they are generally less extincted than the known members.
    \item We found 28 new sources with excess IR emission suggesting the presence of disks.
    \item A proper motion analysis of the $\rho$~Ophiuchi region reveals the presence of two main populations: the first population (Pop 1) of 1022 sources comprises clusters of young stars around the $\rho$~Ophiuchi star and the main Ophiuchus clouds (L1688, L1689, L1709), while the second population (Pop 2) of 304 sources is slightly older and more dispersed, with a similar but distinct proper motion from the first. Both populations occupy approximately the same 3D volume. The second population's age and proper motion suggest that it may have originated from the Upper Sco population. 
    \item The two populations are moving away from each other at about 4.1\,km/s, and will no longer be overlapping in about 4\,Myr. 
    \item Future studies of this benchmark region should treat these two populations separately or risk biasing the star formation observables, such as star formation history, rate, efficiency, or the IMF.
    \item The algorithm used in this paper (OCSVM, \citealt{Ratzenboeck}) has proven to be an effective method for identifying stars belonging to a particular population, based on the properties of a subsample of known sources.
    

\end{enumerate}

\begin{acknowledgements}
We thank the anonymous referee for their thorough and constructive report that has helped us clarify some points and has improved the paper. This project has received funding from the European Research Council (ERC) under the European Union’s Horizon 2020 research and innovation programme (grant agreement No. 851435).
This project has received funding from the Austrian Research Promotion Agency (FFG) under project number 873708.
This research has used data from the European Space Agency (ESA) mission \textit{Gaia} (\url{https://www.cosmos.esa.int/web/gaia}), processed by the Gaia Data Processing and Analysis Consortium (DPAC, \url{https://www.cosmos.esa.int/web/gaia/dpac/consortium}), as well as data from the  European Southern Observatory (ESO) survey \textit{Gaia}-ESO \citep{GES} and the APO Galactic Evolution Experiment \textit{APOGEE} \citep{APOGEE}. This research has also made use of data products from the Two Micron All Sky Survey \textit{2MASS} \citep{2MASS}, which is a joint project of the University of Massachusetts and the Infrared Processing and Analysis Center/California Institute of Technology, funded by the National Aeronautics and Space Administration and the National Science Foundation, the Wide-field Infrared Survey Explorer \textit{WISE} \citep{WISE}, which is a joint project of the University of California, Los Angeles, and the Jet Propulsion Laboratory/California Institute of Technology, funded by the National Aeronautics and Space Administration. The astronomical database \verb|SIMBAD| \citep{SIMBAD} has also greatly contributed to this work. Furthermore, this research was made possible thanks to the use of TOPCAT \citep{Topcat}, an interactive graphical viewer and editor for tabular data, and Python (\url{https://www.python.org/}), in particular Astropy \citep{Astropy}, a community-developed core Python package for Astronomy, NumPy \citep{Numpy2011} and Matplotlib \citep{Matplotlib}, as well as the Aladin Sky Atlas, developed at CDS, Strasbourg Observatory \citep{Aladin}.
\end{acknowledgements}

\bibliographystyle{aa}
\bibliography{oph.bib}

\appendix

\section{Training set criteria} \label{app:train}

In this Appendix, we describe the quality cuts determined for the training set, which were used in the classification algorithm to identify new members. We used the tangential velocities $v_{\alpha}$ and $v_{\delta}$ and their errors for determining the cuts of the training set. The tangential velocities were calculated through the parallaxes $\varpi$ and proper motions $\mu_{\alpha}^{*}$ and $\mu_{\delta}$ using the following formulas:
\begin{align} \label{equ:valpha}
    &v_{\alpha} = 4.74047 \cdot \mu_{\alpha}^{*} / \varpi , \\
    &v_{\alpha\_\text{err}} = 4.74047 \cdot \sqrt{\mu_{\alpha\_\text{err}}^{*2}/\varpi^2+ \mu_{\alpha}^{*2} \cdot \varpi_{\text{err}}^2 / \varpi^4} , \\
    &v_{\delta} = 4.74047 \cdot \mu_{\delta} / \varpi , \\
    &v_{\delta\_\text{err}} = 4.74047 \cdot \sqrt{\mu_{\delta\_\text{err}}^{*2}/\varpi^2+ \mu_{\delta}^{*2} \cdot \varpi_{\text{err}}^2 / \varpi^4} . \label{equ:vdelta}
\end{align}

\begin{figure*}[t!]
    \centering
    \includegraphics[width=\textwidth]{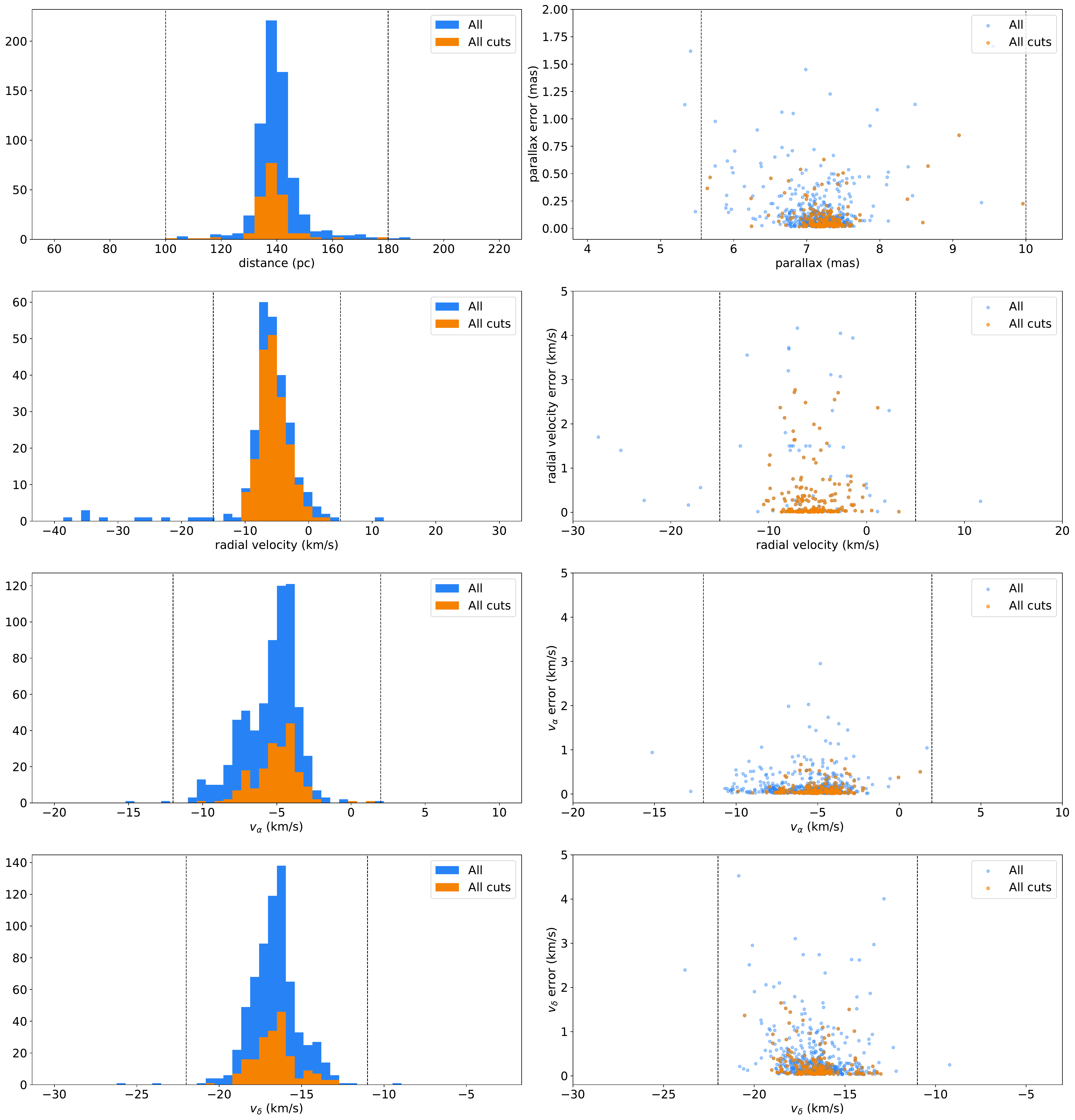}
    \caption{Plots of various properties of the $\rho$~Oph literature sample shown in blue, while the sources that fulfill all of the quality cuts are shown in orange. These plots were used as a visual aid to determine the cuts for the training set.}
    \label{fig:allcuts}
\end{figure*}

The cuts for the training set were determined by using plots as a visual aid. Figure~\ref{fig:allcuts} shows plots of various properties of the complete literature sample in blue and sources that satisfy our chosen quality cuts in orange. We applied the following quality cuts for constructing the training set:
\begin{flalign}
&100\, \text{pc} < d < 180\, \text{pc} ,\\
&\varpi_{\text{err}}/\varpi  < 0.2 ,\\
& -15\, \text{km/s} < v_r < 5\, \text{km/s} ,\\
&v_{r\_\text{err}} < 3 \,\text{km/s} ,\\
&-12 \,\text{km/s} < v_{\alpha} < 2\, \text{km/s} ,\\
&v_{\alpha\_\text{err}} < 3\, \text{km/s} ,\\
&-22\, \text{km/s} < v_{\delta} < -11\, \text{km/s} ,\\
&v_{\delta\_\text{err}} < 3\, \text{km/s}.
\end{flalign}

As the sources are located around a distance $d$ of 140 pc, we applied a symmetrical distance range of 100 to 180 pc for the training set. A relative error-to-value cut was also applied for the parallax $\varpi$. Radial velocities $v_r$ are mostly around a value of -5 km/s, so we applied a symmetrical range of -15 to 5 km/s. A relative error cut is not sensible for the radial velocities since many of them are close to zero, which could lead to losing sources that actually belong to $\rho$~Oph. Therefore, we applied an absolute radial velocity error cut. Since the errors of the tangential velocities $v_{\alpha}$ and $v_{\delta}$ are comparable to the radial velocity errors, similar cuts can be made in all three velocity directions. We applied the same absolute error cut to the tangential velocities, since several $v_{\alpha}$ values are also close to zero. These conditions select sources that do not deviate much from the average values of the chosen properties, creating a suitable selection for finding new sources with similar properties.

\section{\textit{Gaia} quality criteria} \label{app:Gaia}

For the observational HRD in Figure~\ref{fig:cmd}, we applied quality cuts to \textit{Gaia} sources in order to reduce contamination by inferior data, similar to the cuts used in \cite{Grossschedl2020}. Further details on the \textit{Gaia} parameters can be found on the official website of the mission: \url{https://gea.esac.esa.int/archive/documentation/index.html}. We applied the following quality criteria to \textit{Gaia} sources:
\begin{flalign}
&\varpi_{\text{err}}/\varpi < 0.2, \\
&\text{ruwe} < 1.4, \\
&G_{\text{err}} < 0.05 \text{mag}, \\
&\text{visibility\_periods\_used} > 6, \\
&\text{astrometric\_sigma5d\_max} < 1.4.
\end{flalign}

\noindent The $G_{\text{err}}$ value is defined as:
\begin{equation}
  G_{\text{err}} = 1.0857\cdot \text{phot\_g\_mean\_flux\_error} / \text{phot\_g\_mean\_flux}.
\end{equation}

\section{Contamination fraction constraint}

\begin{figure}[t!]
    \centering
    \includegraphics[width=\columnwidth]{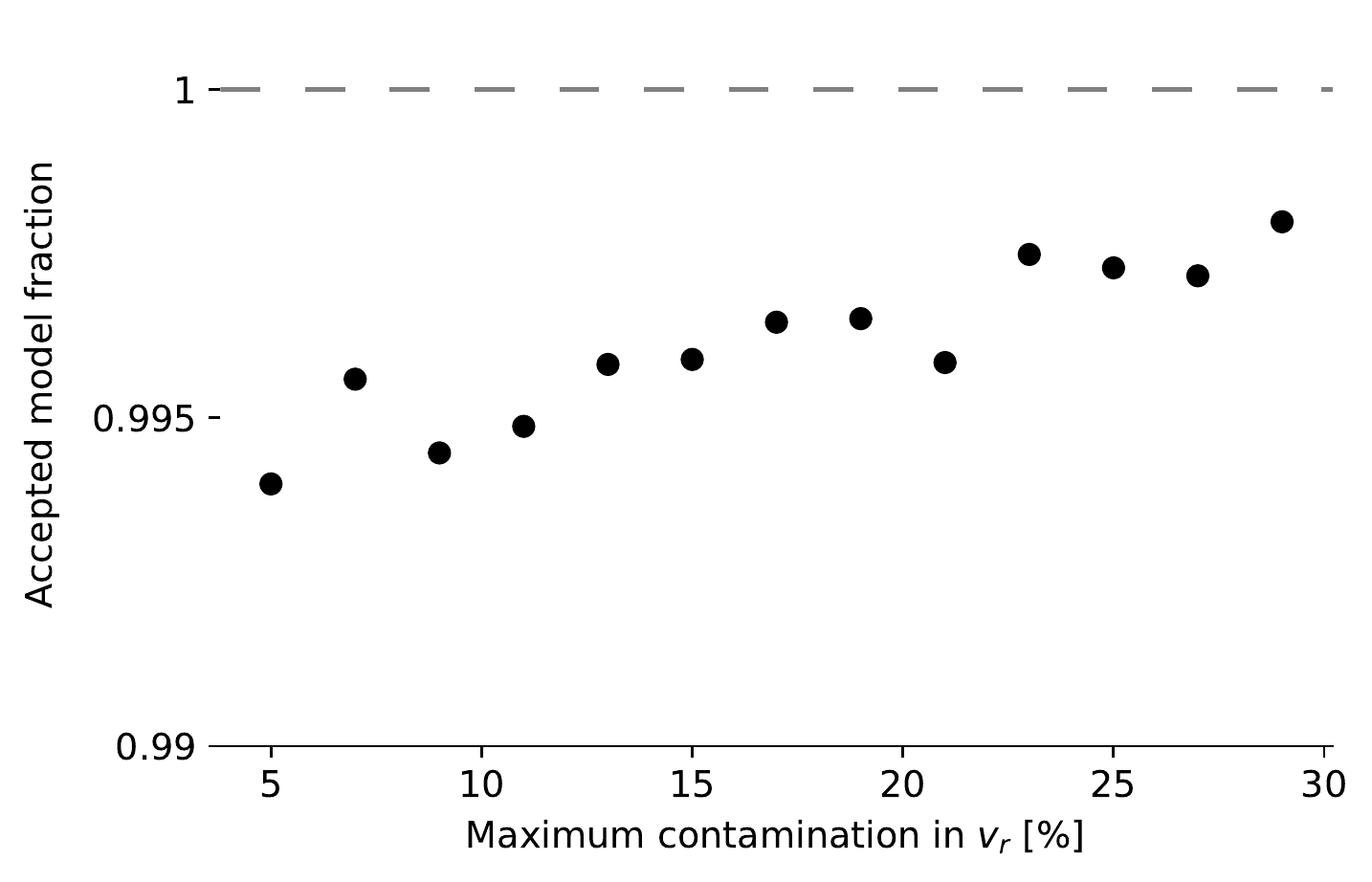}
    \caption{Accepted model fraction according to various maximal contamination requirements. The prior assumption value was varied between $5\,\%$ and $30\,\%$. We found no significant impact of the contamination fraction restriction for individual models on the number of accepted models.}
    \label{fig:vr_contamination}
\end{figure}

Following \cite{Ratzenboeck}, we seek to constrain the contamination fraction of predicted sources across models. As discussed in Sect.~\ref{sec:methods}, the contamination fraction is determined via the 3D velocity distribution of $\rho$~Oph candidate sources. However, for single models, we observed few sources that feature radial velocity measurements in the prediction set, which leads to a marginal effect of the contamination fraction prior assumption on the number of rejected models. This effect is highlighted in Figure~\ref{fig:vr_contamination}, where we see that over $99\,\%$ of models adhere to the contamination rule across various maximal threshold values. For each contamination threshold value we sampled 20 models where we have set the maximal number of samples to $800$ and sampled the remaining prior assumptions within their respective ranges (see Sect.~\ref{sec:methods} for more details). The reported accepted model fraction constitutes a mean value across the 20 sampled prior assumption tuples. The standard deviation is negligibly small. 

\section{Sampling in prior assumption space}

\begin{figure*}[t!]
    \centering
    \includegraphics[width=\textwidth]{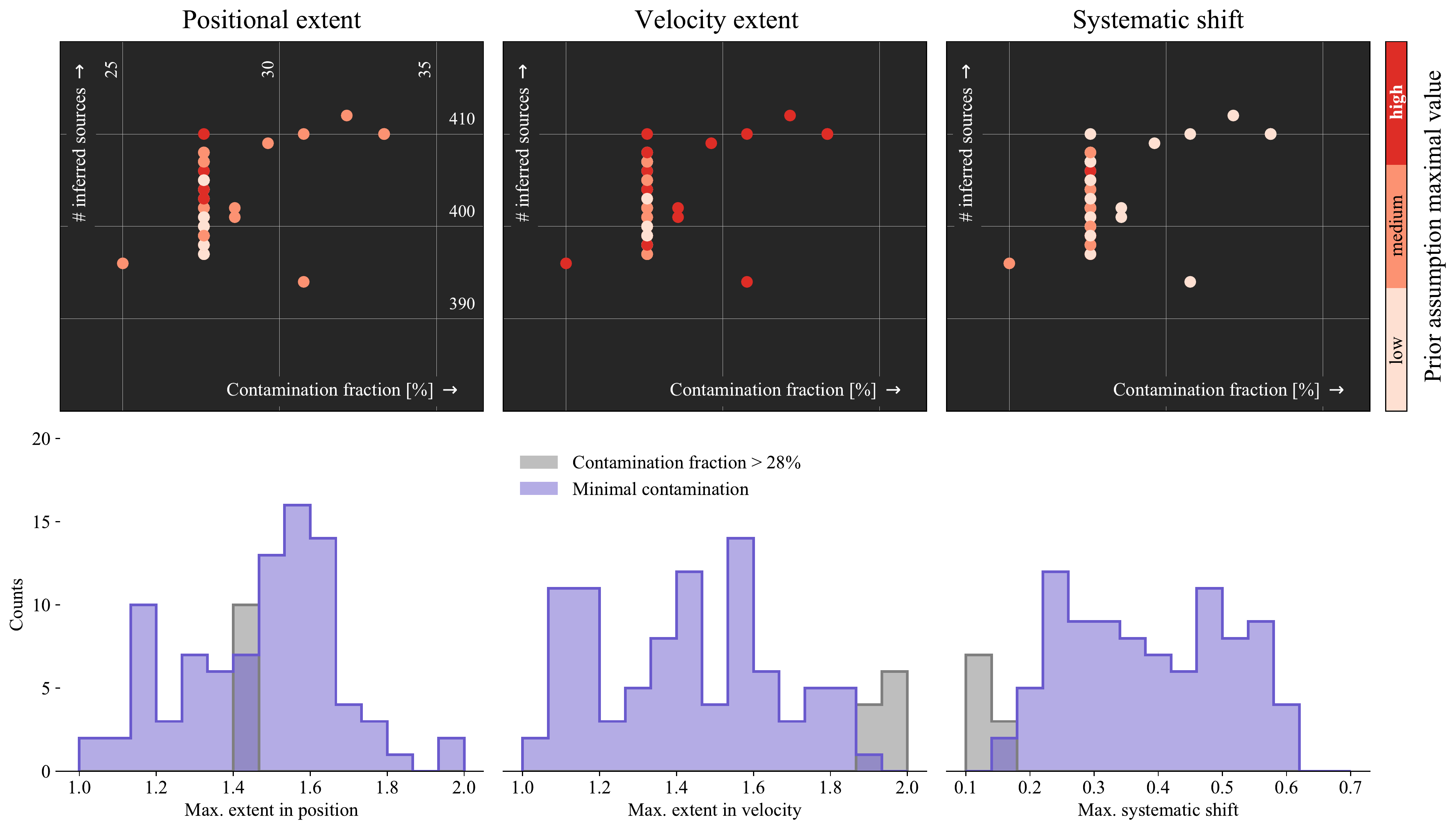}
    \caption{Distribution of the number of predicted sources and contamination fraction space of these ensemble classifiers. \emph{Top.} Distribution of 100 ensemble classifiers trained using various prior assumption constraints in the number of predicted sources and contamination fraction space. We have randomly sampled the prior assumption of the maximal positional extent (left column), the maximal velocity extent (middle column) and the maximal systematic shift (right column) within their respective ranges. The color highlights the maximal prior assumption value. 
    \emph{Bottom.} Sampled prior assumption distributions for models showing a contamination of less than $0.28$ (in purple) and remaining models (in gray). For models with a higher contamination fraction we observe a tendency to higher velocity dispersion and a small systematic shift.}
    \label{fig:pa_sampled}
\end{figure*}

Following the discussion in Sect.~\ref{sec:methods}, we randomly sampled 100 prior assumption tuples within their respective range, which resulted in 100 model ensembles. In Figure~\ref{fig:pa_sampled} the distribution of the number of predicted sources and contamination fraction space of these ensemble classifiers is shown. The prior assumption space of the maximal positional extent (left column), the maximal velocity extent (middle column) and the maximal systematic shift (right column) was uniformly sampled within their respective ranges. We use color to encode the maximal prior assumption value in this space. On the bottom, the sampled prior assumption distributions for models showing minimal contamination (in purple) and the remaining models (in gray) can be seen. In models with high contamination, we observe a tendency to higher velocity dispersion but low systematic shifts. We observe that ``good'' models with lower contamination experience sometimes even a drastic systematic shit. This shift is due to the second population we uncovered.

\section{Stability} \label{app:stab}

\begin{figure}[t!]
    \centering
    \includegraphics[width=\columnwidth]{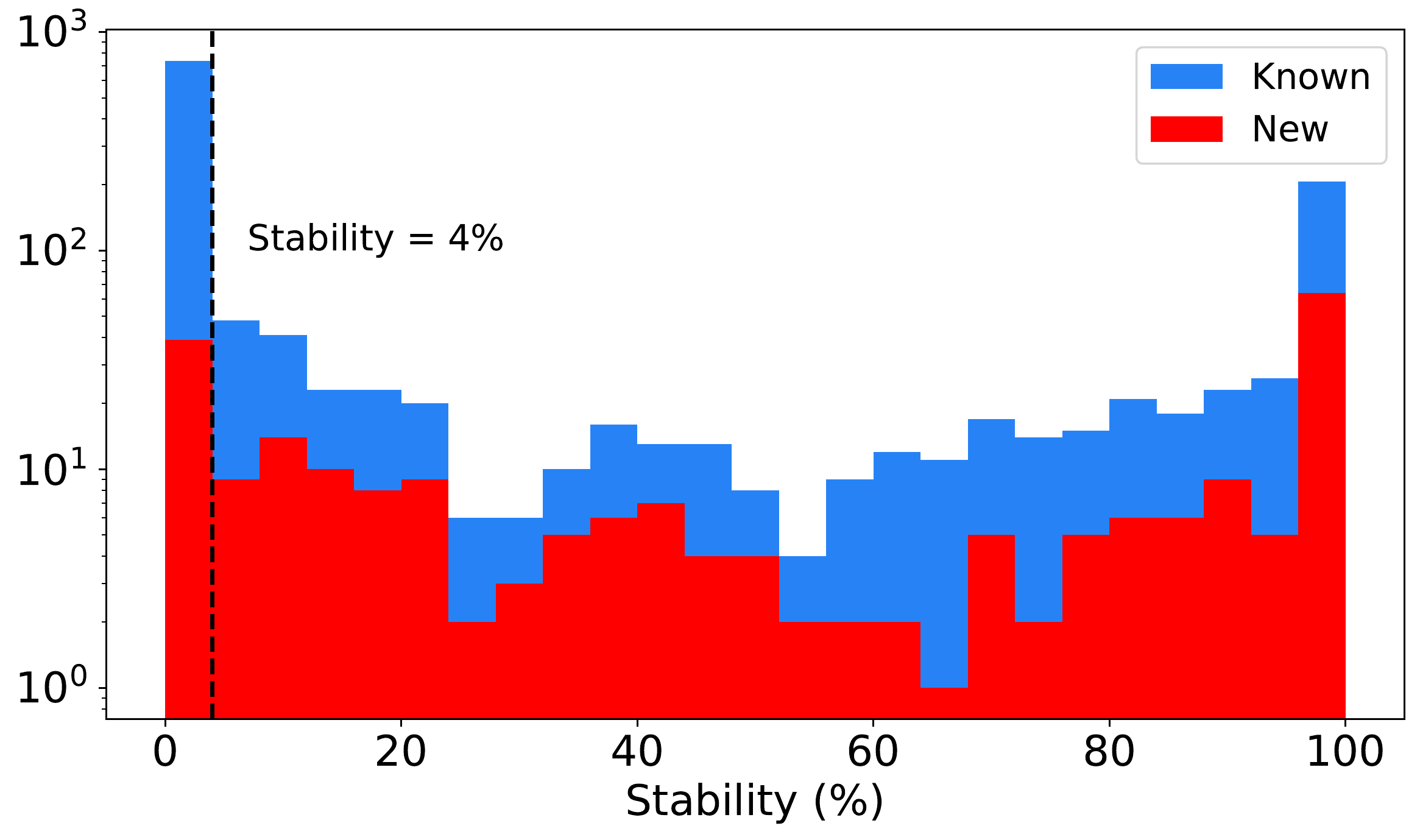}
    \caption{The stability (in percent) of the known and new sources, as determined by the OCSVM method.}
    \label{fig:stab}
\end{figure}

We discuss the stability of the predicted sources as well as the stability cut we chose. Although the model selection process via a set of prior assumptions (see Sect.~\ref{sec:methods}) removed a majority of unsuitable models, the lack of a clear objective function still leaves some contamination in our final prediction sample. To find a set of high-fidelity members, we studied the prediction frequency, or stability, of the predicted sources across the model ensemble. Figure~\ref{fig:stab} shows a histogram of the stability of the known and new sources. Both of them show a relatively similar stability distribution. Many of the known sources from the literature are predicted with a stability of 0 because they are not in the \textit{Gaia} EDR3 catalog.

As discussed in \cite{Ratzenboeck}, an appropriate stability threshold should reduce spurious sources while maximizing the number of legitimate cluster members. For this purpose, the authors studied the impact of the stability criterion on the Cartesian velocity dispersion and selected an optimal value by eye. Now we aimed to train multiple model ensembles under different prior assumptions and jointly attempt to characterize each model ensemble corresponding to a single prior belief tuple in terms of a contamination estimate and the number of identified points at their respective optimal stability thresholds (see Sect.~\ref{sec:methods}). Therefore, we intend to automatically determine a threshold value for each model ensemble. To do so we considered the following. The distribution of predicted members and training members in 5D is by design very close and adheres to our prior assumptions, so we cannot infer an independent quality criterion from the prediction in 5D. However, since stars that are born together move together \citep{Kamdar2019}, we can, similarly to \cite{Ratzenboeck}, use the, albeit sparsely available, full 3D velocity information for determining the stability criterion. 

To be co-moving, we postulate that the predicted sources with radial velocity information should be distributed as similarly as possible to the training set 3D velocities. To test this similarity, we modeled the 3D velocity data using a multivariate normal distribution. We determined the mean and covariance by maximizing the likelihood of the training data under the model. To estimate the difference between the trained and predicted sources, we used the  Kullback–Leibler (KL) divergence \citep{kullback1951} D$_{\text{KL}} (p~\lVert~q)$ where $q$ and $p$ both constitute probability distribution functions. The KL divergence of $p(\boldsymbol{x})$ from $q(\boldsymbol{x})$ of the continuous variable $\boldsymbol{x}$ is defined via
\begin{equation}
    \text{D}_{\text{KL}} (p~\lVert~q) = \int_{-\infty}^{\infty} p(\boldsymbol{x})~ \text{log} \left( \frac{p(\boldsymbol{x})}{q(\boldsymbol{x})} \right).
\end{equation}
It can be interpreted as the information content that is lost when the true distribution $p$ is substituted by an approximate distribution $q$ \citep{Burnham2002}. Here, $p$ represents our training set distribution, while the approximate distribution $q$ describes the distribution of predicted sources. To evaluate D$_{\text{KL}} (p~\lVert~q)$, we modeled $q$, the velocity distribution of the derived members, assuming a single Gaussian. For two multivariate normal distributions, the KL divergence can be written analytically in the following form \citep{MatrixCook2012}:

\begin{equation}
    \text{D}_{\text{KL}} = \frac{1}{2} \left[ \text{log} \frac{|\Sigma_q|}{|\Sigma_p|} - d + \text{tr}(\Sigma_q^{-1}\Sigma_p) + (\mu_q - \mu_p)^{T} \Sigma_q^{-1} (\mu_q - \mu_p) \right].
\end{equation}

\begin{figure}[t!]
    \centering
    \includegraphics[width=\columnwidth]{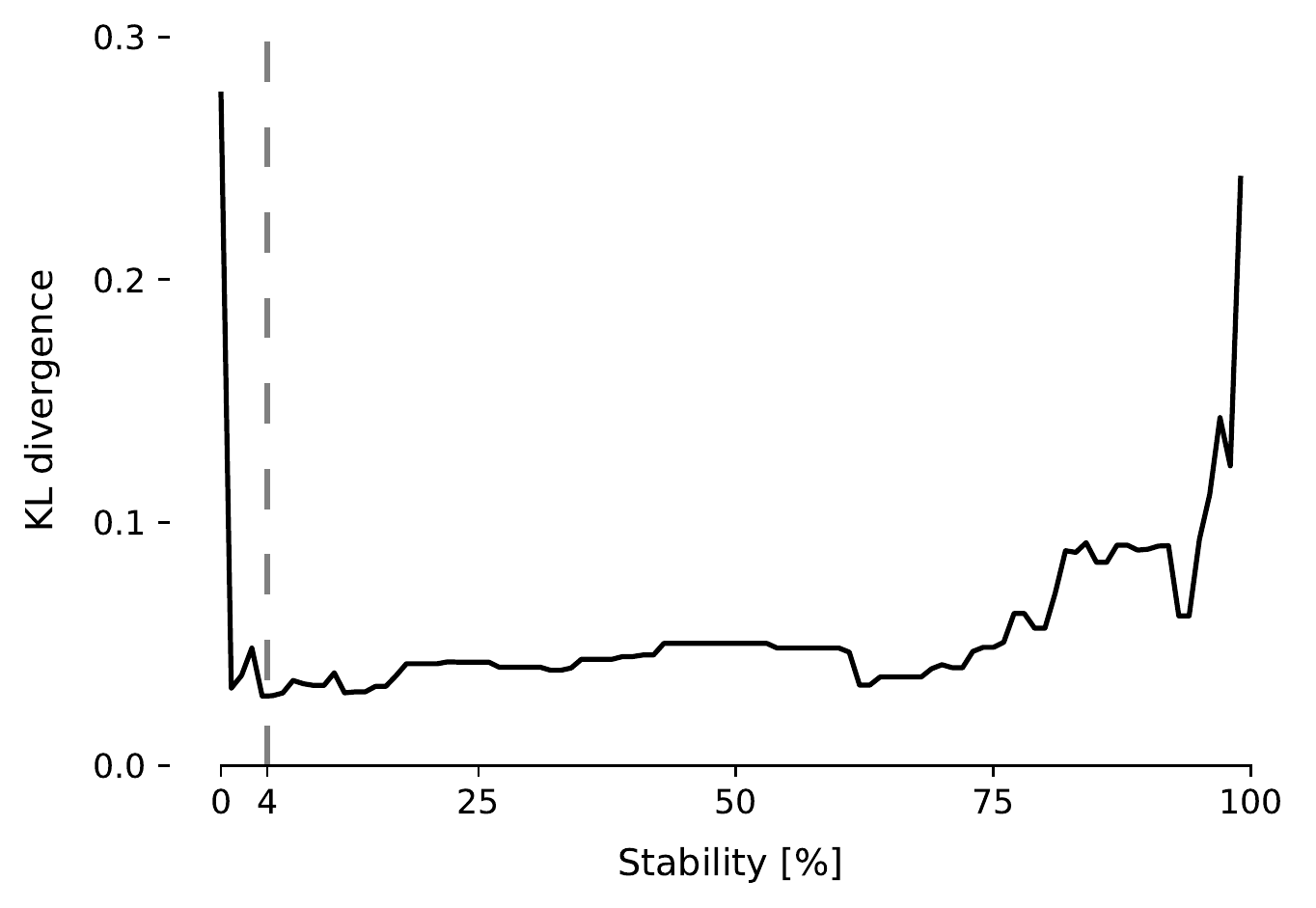}
    \caption{KL divergence between Cartesian velocity distributions of training and predicted source populations determined across various stability threshold values. We found an optimal threshold criterion of stability > $4\,\%$ for the final ensemble across various prior assumptions that produce minimal contamination (see Sect.~\ref{sec:methods} for a more detailed discussion).}
    \label{fig:stab_kldiv}
\end{figure}

Here, $\mu$ and $\Sigma$ refer to the mean and covariance matrices of the multivariate normal distributions, respectively. The variable $d$ describes the number of dimensions, which is in this case $d=3$. 
To find the optimal stability threshold we seek to minimize the KL divergence between the Cartesian velocity distribution of training and predicted sample populations, which is illustrated in Figure~\ref{fig:stab_kldiv}. We found an optimal threshold criterion of stability > $4\,\%$. 
The stability is included in our final catalog shown in Table~\ref{tab:ophtab}.

\section{Validation of predicted sources in the HRD}

\begin{figure}[t!]
    \centering
    \includegraphics[width=\columnwidth]{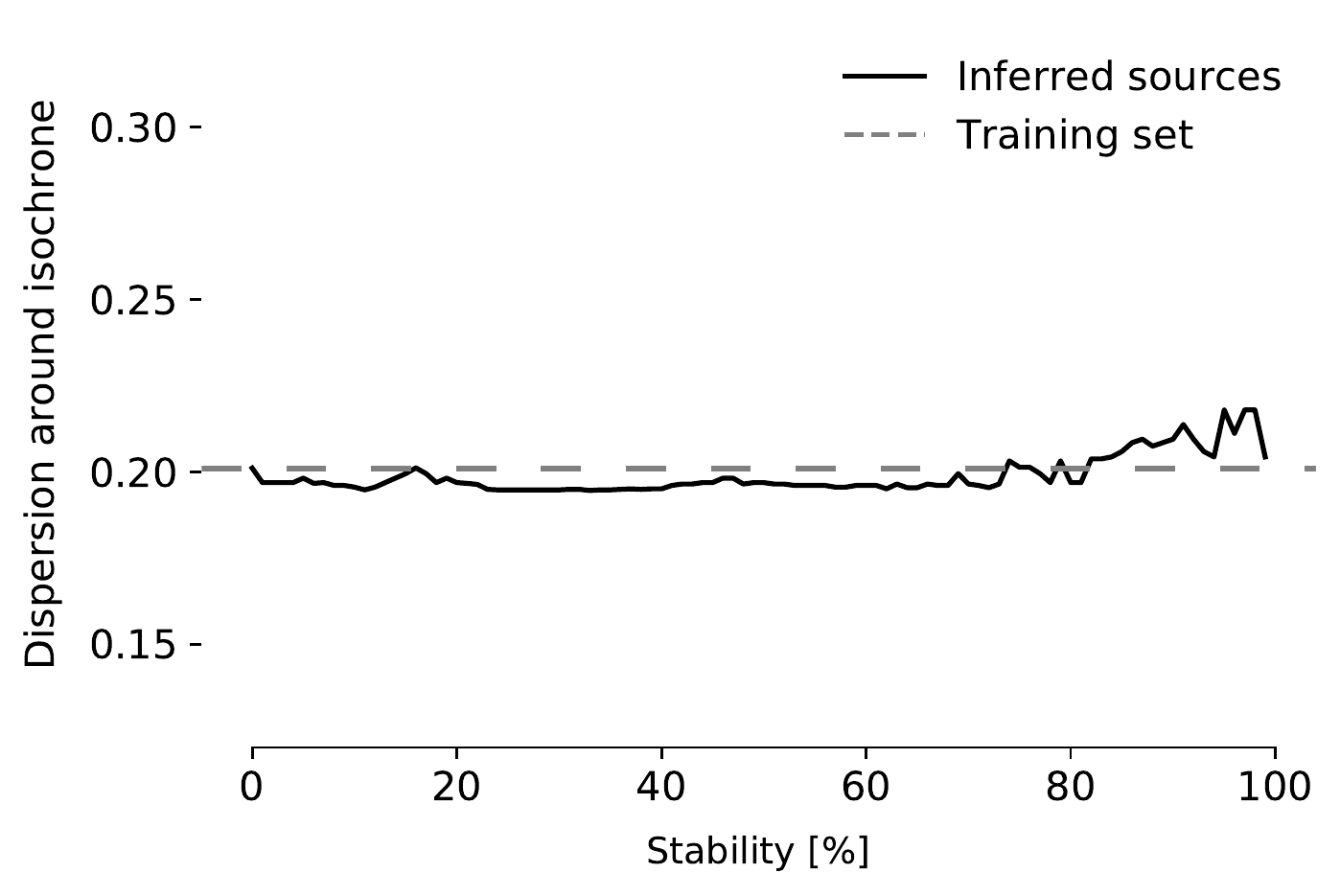}
    \caption{Comparison between the training and predicted (inferred) sources across the full stability range. The y-axis shows the standard deviation of residuals between the data and an isochrone of $5\,$Myr, describing the best fit to the training data. We find no significant difference between the training set members and the predicted sources based on their HRD distributions.}
    \label{fig:stab_isochrone}
\end{figure}

As a final validation step, we compare the predicted source distribution to the training set distribution in the HRD. Since both populations should be coeval, we can characterize the HRD distribution by their deviation from the best fitting isochrone on the training set. In Figure~\ref{fig:stab_isochrone}, the standard deviation of residuals between the data and the $5\,$Myr isochronal curve is shown. We found no significant difference between the training set members and the predicted sources based on their HRD distributions.

\section{Astrometric properties of known and new sources}

Table \ref{tab:avg} shows the average astrometric properties of the sources in $\rho$~Oph, such as the distances, proper motions, radial and tangential velocities, Galactic Cartesian positions $X,Y,Z$ and Galactic Cartesian velocities $U,V,W$, as well as the standard deviations (1$\sigma$) of these parameters. These average values were determined for the known and new sources, as well as for all of them together. To avoid the influence of outliers, impostors defined in Sect.~\ref{sec:imp} were not included in the calculations. The column $\Delta$ contains the difference of the known and new mean values for comparison. 

There appear to be only small deviations between the properties of the known and new sources, which are not significant within $1 \sigma$. This further confirms that, on average, they belong to the same region. The average values of $\rho$~Oph for $\varpi, \mu_{\alpha}^{*}, \mu_{\delta}, X, Y, Z$ agree relatively well with those determined by \cite{Canovas} within 1$\sigma$.

\begin{table}[t!]
    \centering
    \caption{Average astrometric properties of the known and new sources in $\rho$~Oph.}
    \begin{tabular}{lrrrr}
    \hline
    \hline
    Dimension & Known & New & All & $\Delta$ \\
    \hline
$\alpha$ (deg) & 246.6$\pm$1.0 & 245.2$\pm$1.6 & 246.3$\pm$1.3 & 1.4 \\
$\delta$ (deg) & -24.2$\pm$0.8 & -23.8$\pm$1.5 & -24.1$\pm$0.9 & -0.4 \\
$\varpi$ (mas) & 7.2$\pm$0.4 & 7.0$\pm$0.3 & 7.1$\pm$0.4 & 0.1 \\
$d$ (pc) & 140.0$\pm$8.5 & 142.5$\pm$5.8 & 140.6$\pm$7.9 & -2.4 \\
$\mu_{\alpha}^{*}$ (mas/yr) & -8.2$\pm$3.0 & -9.6$\pm$1.9 & -8.6$\pm$2.8 & 1.4 \\
$\mu_{\delta}$ (mas/yr) & -24.9$\pm$2.5 & -23.9$\pm$2.0 & -24.7$\pm$2.4 & -1.0 \\
$v_{\alpha}$ (km/s) & -5.4$\pm$1.9 & -6.5$\pm$1.3 & -5.7$\pm$1.8 & 1.1 \\
$v_{\delta}$ (km/s) & -16.7$\pm$1.6 & -16.1$\pm$1.1 & -16.5$\pm$1.5 & -0.6 \\
$v_r$ (km/s) & -5.8$\pm$4.3 & -5.0$\pm$5.1 & -0.9$\pm$68.5 & -0.8 \\
$X$ (pc) & 132.7$\pm$8.1 & 134.1$\pm$5.8 & 133.0$\pm$7.6 & -1.4 \\
$Y$ (pc) & -15.8$\pm$1.9 & -17.4$\pm$3.0 & -16.2$\pm$2.4 & 1.6 \\
$Z$ (pc) & 41.8$\pm$3.4 & 44.6$\pm$4.2 & 42.5$\pm$3.9 & -2.9 \\
$U$ (km/s) & -5.2$\pm$3.3 & -4.9$\pm$4.5 & -5.9$\pm$5.8 & -0.2 \\
$V$ (km/s) & -15.3$\pm$1.4 & -15.9$\pm$1.0 & -15.3$\pm$1.5 & 0.6 \\
$W$ (km/s) & -8.6$\pm$2.1 & -7.3$\pm$2.5 & -8.7$\pm$2.5 & -1.3 \\
    \hline
    \end{tabular}
    \tablefoot{The average positional and dynamical values, including their standard deviations (1$\sigma$), were determined for the known and new sources separately, as well as for all of them together. The column $\Delta$ contains the difference of the known and new mean values for comparison of the two.}
    \label{tab:avg}
\end{table}

\section{$\rho$~Oph catalog overview} \label{app:tab}

In this Appendix we present our final catalog of $\rho$~Oph sources, which is available online at the CDS. It includes all known sources from the literature and all sources identified by the OCSVM, even those with a stability < 4, resulting in a total of 1343 sources. Table~\ref{tab:ophtab} shows an overview of the column names, their units and their descriptions. In total, our catalog contains 67 columns. The column ``Ref'' serves as a reference for the literature sources, where each paper is cited by their reference number given in Table~\ref{tab:literature}. Several sources were obtained from more than one paper; therefore, some sources have more than one reference number.

Since the known sources have proper motions and radial velocities obtained from the literature, \textit{Gaia} EDR3, APOGEE-2, or \textit{Gaia}-ESO, we provide the column ``Ref\_pm\_rv'' for the reference of the proper motions and radial velocity values, respectively. Each row contains two numbers for citation of these values, where ``1,'' ``2,'' ``3,'' and ``4'' signify measurements obtained from the literature, \textit{Gaia} EDR3, APOGEE, and \textit{Gaia}-ESO, respectively. ``0'' implies that a source does not have a corresponding proper motion, parallax or radial velocity measurement.

\begin{center}
\onecolumn
\begin{longtable}{llp{9cm}}
\caption{Column overview of the final catalog containing known and new $\rho$~Oph sources.}  \label{tab:ophtab} \\

\hline \hline \multicolumn{1}{l}{Column Name} & \multicolumn{1}{l}{Unit} & \multicolumn{1}{l}{Description} \\ \hline 
\endfirsthead
\multicolumn{3}{c}%
{{\bf{\tablename\ } \thetable{} -- continued from previous page}} \\
\hline \hline \multicolumn{1}{l}{Column Name} & \multicolumn{1}{l}{Unit} & \multicolumn{1}{l}{Description} \\ \hline 
\endhead

\hline \multicolumn{3}{r}{{Continued on next page}} \\ \hline
\endfoot
\hline \hline
\endlastfoot

        source\_id\_edr3 & - &\textit{Gaia} EDR3 ID \\
        ra & deg& Right ascension (J2000) \\
        dec& deg & Declination (J2000) \\
        l & deg & Galactic longitude \\
        b & deg & Galactic latitude \\
        parallax & mas & Parallax \\
        parallax\_error & mas& Parallax error \\
        distance & pc & Distance, determined from the inverse of the parallax \\
        pmra & mas/yr & Proper motion in ra direction \\
        pmra\_error & mas/yr & Error in pmra \\
        pmdec & mas/yr & Proper motion in dec direction \\
        pmdec\_error & mas/yr & Error in pmdec \\
        radial\_velocity & km/s & Heliocentric radial velocity \\
        radial\_velocity\_error & km/s & Error in radial velocity \\
        v\_alpha & km/s & Tangential velocity in ra direction \\
        v\_alpha\_error & km/s & Error in v\_alpha \\
        v\_delta & km/s & Tangential velocity in dec direction \\
        v\_delta\_error & km/s & Error in v\_delta \\
        X & pc& Galactic Cartesian X position component\\
        Y & pc& Galactic Cartesian Y position component \\
        Z & pc& Galactic Cartesian Z position component\\
        U & km/s& Galactic Cartesian U velocity component\\
        V & km/s& Galactic Cartesian V velocity component \\
        W & km/s& Galactic Cartesian W velocity component\\
        ruwe & - & Renormalized unit weight error  \\
        astrometric\_sigma5d\_max & mas & Longest principal axis in the 5D error ellipsoid \\
        astrometric\_params\_solved & - & Which parameters have been solved for \\
        visibility\_periods\_used & - & Number of visibility periods in the astrometric solution \\
        phot\_g\_mean\_flux & e-/s & G-band mean flux \\
        phot\_g\_mean\_flux\_error & e-/s & Error on G-band mean flux \\
        phot\_g\_mean\_mag & mag & G-band mean magnitude \\
        phot\_bp\_mean\_mag & mag & Integrated BP mean magnitude \\
        phot\_rp\_mean\_mag & mag & Integrated RP mean magnitude \\
        bp\_rp & mag & BP--RP color \\
        Train & -& =1 for sources in the training set \\
        Predict & -& =1 for predicted sources in \textit{Gaia} EDR3 \\
        New & -& =1 for new sources in \textit{Gaia} EDR3 \\
        Stability & -& Stability of the sources, range: 0--100 \\
        Impostors & - & =1 for impostor sources \\
        pml & mas/yr & Proper motion in l direction\\
        pmb & mas/yr & Proper motion in b direction\\
        v\_l & km/s & Velocity in l direction \\
        v\_b & km/s & Velocity in b direction \\
        angle\_l\_hel & deg & Heliocentric proper motion angle to l-axis \\
        pml\_lsr & mas/yr & Proper motion in l direction (LSR) \\
        pmb\_lsr & mas/yr & Proper motion in b direction (LSR) \\
        v\_l\_lsr & km/s & Velocity in l direction (LSR) \\
        v\_b\_lsr & km/s & Velocity in b direction (LSR) \\
        angle\_l\_lsr & deg & LSR proper motion angle to l-axis \\
        Pop & - & =1 for Pop~1 sources, =2 for Pop~2 sources, =0 if neither \\
        IR\_excess & - & =1 for YSOs with IR excess, =0 for Class\,III sources \\
        designation\_2MASS & - & \textit{2MASS} ID \\
        j\_m & mag & J-band magnitude \\
        j\_cmsig & mag & Uncertainty in J-band magnitude \\
        h\_m & mag & H-band magnitude \\
        h\_cmsig & mag & Uncertainty in H-band magnitude \\
        k\_m & mag & K-band magnitude \\
        k\_cmsig & mag & Uncertainty in K-band magnitude \\
        designation\_WISE & - & \textit{WISE} ID \\
        w1mpro & mag & \textit{WISE} W1 magnitude\\
        w1snr & -& W1 signal-to-noise ratio \\
        w2mpro & mag & \textit{WISE} W2 magnitude\\
        w2snr & -& W2 signal-to-noise ratio \\
        w3mpro & mag & \textit{WISE} W3 magnitude\\
        w3snr & -& W3 signal-to-noise ratio \\
        Ref & - & Reference for literature sources, see Table \ref{tab:literature}, range: 1--11  \\
        Ref\_pm\_rv & - & Reference for proper motions and radial velocity: literature=1, \textit{Gaia} EDR3=2, APOGEE=3, \textit{Gaia}-ESO=4\\
\end{longtable}
    \tablefoot{Column overview of the final catalog of $\rho$~Oph sources, which includes the known sources from the literature as well as the new sources identified by the algorithm. The complete table is available for download at the CDS.}
\twocolumn
\end{center}

\end{document}